\documentclass[useAMS,usenatbib]{mn2e}

 \usepackage{times}

\usepackage{graphicx}
\usepackage{supertabular}
\usepackage{subfig}
\usepackage{rotating}
\usepackage{lscape}
\usepackage{epsfig}
\usepackage{amssymb}
\usepackage{myaasmacros}
\usepackage{amsmath}
\usepackage{multicol}
\usepackage{slashbox}
\usepackage{color}

\def\lesssim{\mathrel{\hbox{\rlap{\hbox{\lower4pt\hbox{$\sim$}}}\hbox{$<$}}}}
\def\gtrsim{\mathrel{\hbox{\rlap{\hbox{\lower4pt\hbox{$\sim$}}}\hbox{$>$}}}}

\def\arcsec{\hbox{$^{\prime\prime}$}\,}
\def\arcmin{$^{\prime}$\,}
\def\deg{\hbox{$^\circ$}}
\def\mic{$\,\mu $m\,}
\title[The H-ATLAS SDP catalogue]{{\it Herschel}--ATLAS: First data release of the Science Demonstration Phase source catalogues}
\author[E.E. Rigby]{\parbox{\textwidth}{E.E. Rigby$^{1}$\thanks{E-mail:
    emma.rigby@nottingham.ac.uk; emmaerigby@gmail.com}, S.J. Maddox$^{1}$,
  L. Dunne$^{1}$, 
M. Negrello$^{2}$,
D.J.B. Smith$^{1}$, 
J. Gonz{\'a}lez-Nuevo$^{3}$,
D. Herranz$^{4}$,
M. L{\'o}pez-Caniego$^{4}$,
R. Auld$^{5}$, 
S. Buttiglione$^{6}$, 
M.Baes$^{11}$,
A. Cava$^{7}$, 
A. Cooray$^{8}$,
D. L. Clements$^{9}$,
A. Dariush$^{5}$, 
G. De Zotti$^{3,6}$,
S. Dye$^{5}$,
S. Eales$^{5}$, 
D. Frayer$^{10}$,
J. Fritz$^{11}$, 
R. Hopwood$^{2}$,
E. Ibar$^{12}$, 
R.J. Ivison$^{12,13}$, 
M. Jarvis$^{14}$,
P. Panuzzo$^{15}$,
E. Pascale$^{5}$, 
M. Pohlen$^{5}$, 
G. Rodighiero$^{6}$, 
S. Serjeant$^{2}$,
P. Temi$^{16}$,
M. A. Thompson$^{14}$}\vspace{0.4cm}\\
$^{1}$School of Physics and Astronomy, University of Nottingham, University Park, Nottingham NG7 2RD, UK\\
$^{2}$Department of Physics and Astronomy, The Open University, Walton Hall, MK7 6AA Milton Keynes, UK\\
$^{3}$SISSA, Via Bonomea 265, I-34136 Trieste, Italy \\
$^{4}$Instituto de F\'{i}sica de Cantabria (CSIC-UC), Avda. los Castros s/n, 39005 Santander, Spain\\
$^{5}$School of Physics and Astronomy, Cardiff University, The Parade, Cardiff, CF24 3AA, UK\\
$^{6}$INAF â Osservatorio Astronomico di Padova,  Vicolo Osservatorio 5, I-35122 Padova, Italy\\
$^{7}$Departamento de Astrof\'{\i}sica, Facultad de CC. F\'{\i}sicas, Universidad Complutense de Madrid, E-28040 Madrid, Spain\\
$^{8}$Department of Physics and Astronomy, University of California, Irvine, CA 92697, USA \\
$^{9}$Astrophysics Group, Imperial College, Prince Consort Road, London SW7 2AZ, UK\\
$^{10}$Infrared Processing and Analysis Center; California Institute of Technology 100-22, Pasadena, CA 91125, USA\\
$^{11}$Sterrenkundig Observatorium, Universiteit Gent, Krijgslaan 281 S9, B-9000 Gent, Belgium\\
$^{12}$UK Astronomy Technology Centre, Royal Observatory Edinburgh, Edinburgh, EH9 3HJ, UK\\
$^{13}$Institute for Astronomy, University of Edinburgh, Royal Observatory, Edinburgh, EH9 3HJ, UK\\
$^{14}$Centre for Astrophysics, Science \& Technology Research Institute, University of Hertfordshire, Hatfield, Herts, AL10 9AB, UK\\
$^{15}$Centre CEA de Saclay (Essonne), Gif-sur-Yvette, 921191 cedex, France\\
$^{16}$Astrophysics Branch, NASA Ames Research Center, Mail Stop 245-6, Moffett Field, CA 94035, USA}

\begin{document}

\date{}

\pagerange{\pageref{firstpage}--\pageref{lastpage}} \pubyear{2002}

\maketitle

\label{firstpage}

\begin{abstract}
The {\it Herschel}--ATLAS is a survey of 550 square degrees with the {\it Herschel}
Space Observatory in five far--infrared and submillimetre bands. The
first data for the survey, observations of a field 4 $\times$ 4 deg$^{2}$ in size,
were taken during the Science Demonstration Phase, and reach a $5\sigma$
noise level of 33.5 mJy/beam at 250\mic. This paper describes the source extraction methods used to create the
corresponding Science Demonstration Phase catalogue, which contains
6876 sources, selected at 250\mic, within $\sim$14 sq. degrees. SPIRE sources are extracted
using a new method specifically developed for {\it Herschel} data;
PACS counterparts of these sources are identified using circular
apertures placed at the SPIRE positions. Aperture flux densities are
measured for sources identified as extended after matching to optical
wavelengths. The reliability of this catalogue is also discussed,
using full simulated maps at the three SPIRE bands. These show that a
significant number of sources at 350 and 500\mic have undergone flux
density enhancements of up to a factor of $\sim$2, due mainly to
source confusion. Correction factors are determined for these
effects. The SDP dataset and corresponding catalogue will be available from \verb1http://www.h-atlas.org/1. 
\end{abstract}


\section{Introduction}

The {\it Herschel} Astrophysical Terahertz Large Area Survey (H-ATLAS) survey is the largest, in time and area, of the
extragalactic Open Time Key Projects to be carried out with the
European Space Agency (ESA) Herschel Space Observatory
\citep{herschel}\footnote{Herschel is an ESA space observatory with
  science instruments provided by European-led Principal Investigator
  consortia and with important participation from NASA.}. When
complete it will cover $\sim$550 square degrees of the sky, in five far--infrared and
submillimetre bands (100, 160, 250, 350 and 500\mic), to a 5$\sigma$ depth of 33 mJy/beam at 250\mic. The predicted number of
sources is $\sim$200,000; of these $\sim$40,000 are expected to lie within $z<0.3$. A full
description of the survey can be found in \citet{eales}.   

This paper presents the 250\mic selected source catalogue created from the initial
H-ATLAS Science Demonstration Phase (SDP) observations. Eight papers
based on this catalogue have already been published in the A\&A {\it
  Herschel} Special Issue ranging from the identification of blazars \citep{gonzalez} and
debris disks \citep{thompson} in the SDP field, to determinations of
the colours \citep{amblard}, source counts \citep{clements},
clustering \citep{wtheta} and 250\mic luminosity function evolution \citep{dye} of the submillimetre
population, as well as the star formation history of quasar host
galaxies \citep{serjeant2} and the dust energy balance of a nearby
spiral galaxy \citep{baes}.  

The layout of the paper is as follows: Section \ref{hers_obs} describes the SDP observations; Section \ref{extract}
describes the source extraction procedure for the five bands; finally,
Section \ref{sims} outlines the simulations used to quantify the
reliability of the catalogue. 
For more details of the SDP data see \citet{spiremaps} and
\citet{pacsmaps} for the SPIRE and PACS data reduction respectively,
and \citet{Smith} for the multiwavelength catalogue matching. 

\begin{figure}
\centering
\subfloat[Original combined map]{\includegraphics[scale=0.15]{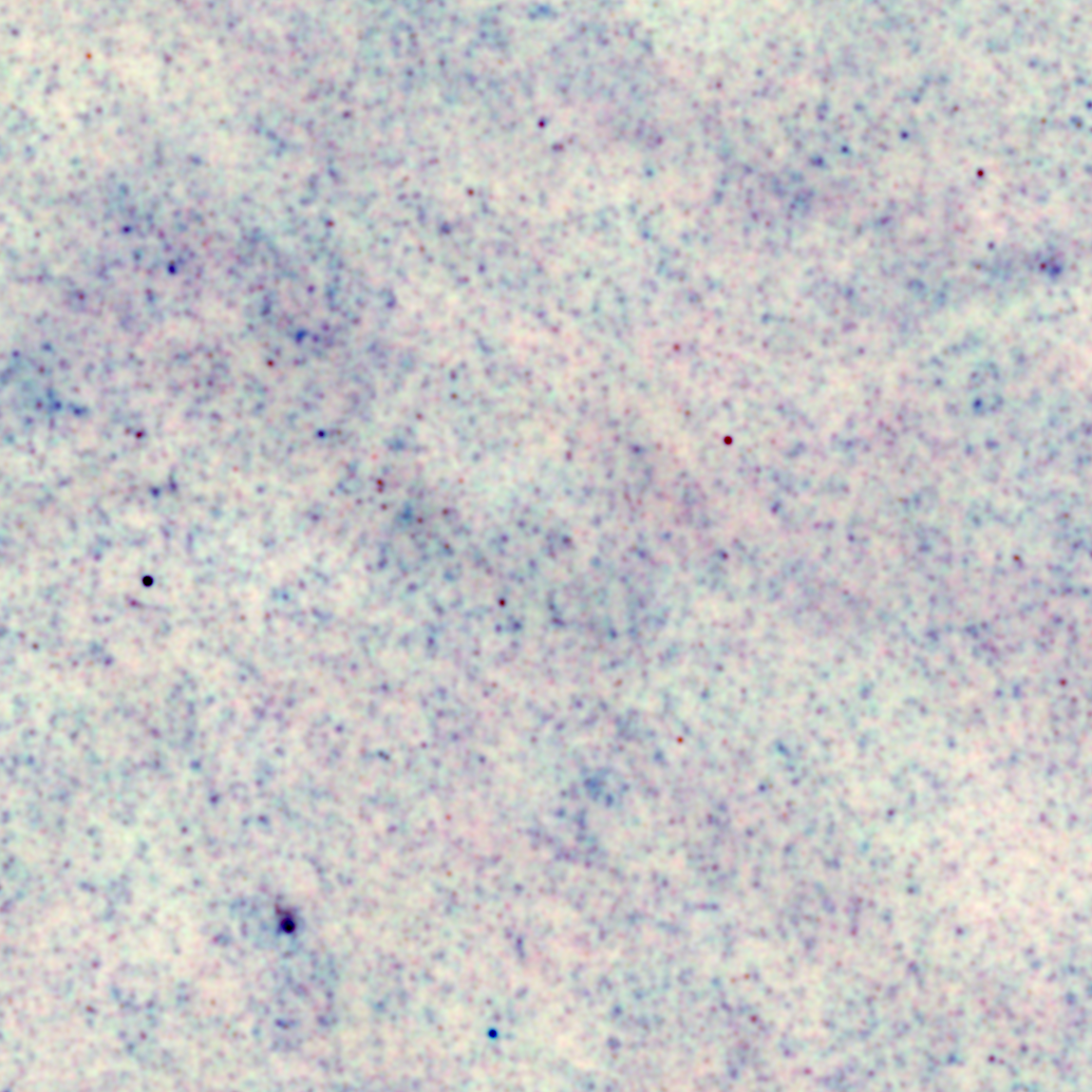}} \\
\subfloat[After background subtraction]{\includegraphics[scale=0.15]{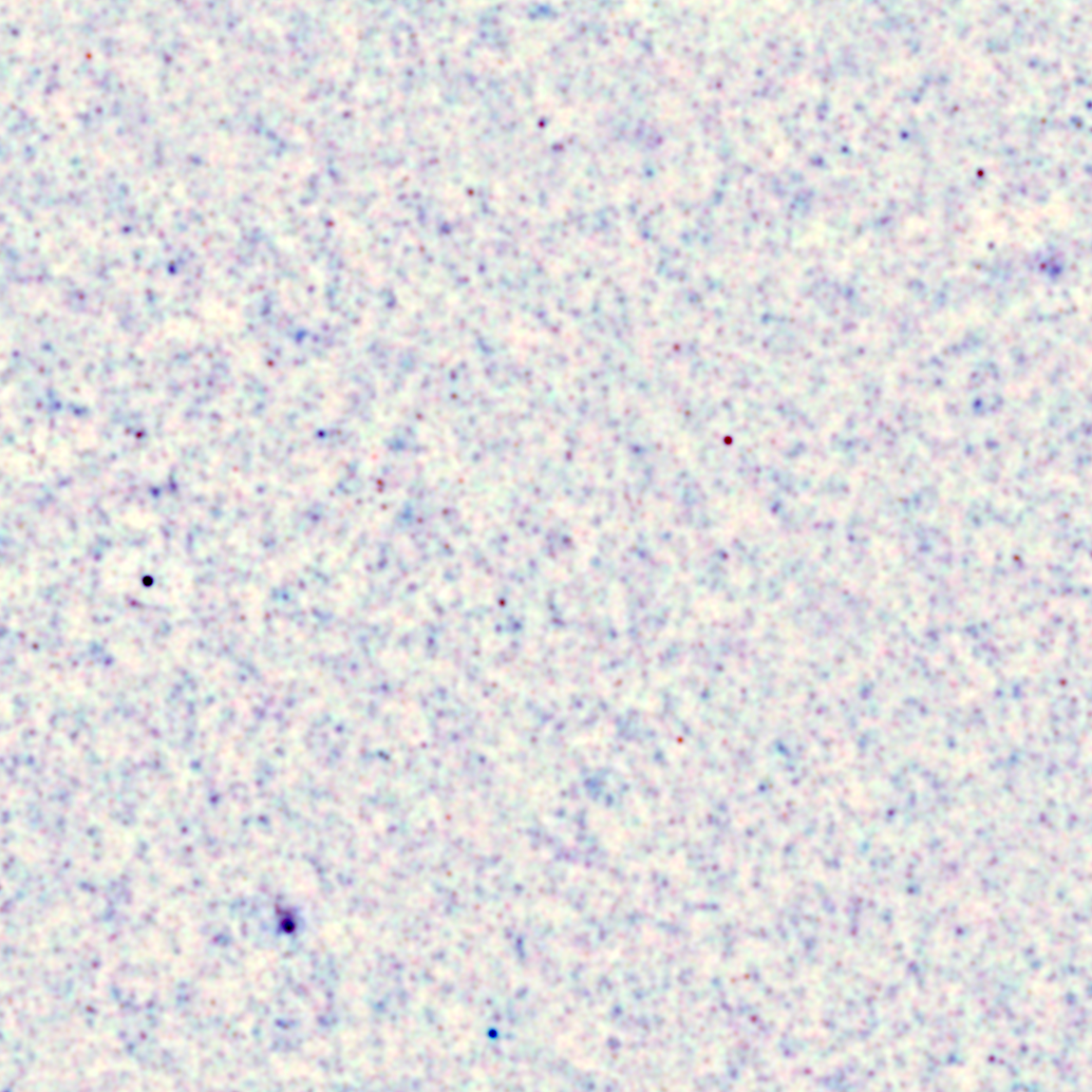}}
\caption{\protect\label{pre_backsub} False--colour images of a 1.5
  sq. degree region of the SDP field showing the three SPIRE bands combined. Image (a) is before
  background--subtraction and shows clear contamination by galactic
  cirrus; image (b) shows the reduction in contamination after subtracting the
  background. }
\end{figure}

\section{{\it Herschel} observations}
\label{hers_obs}

The SDP observations for the H--ATLAS survey cover an area of
$\sim$4\deg$\times$4\deg, centred at
$\alpha$=09$^{h}$05$^{m}$30.0$^{s}$,
$\delta=$00\deg30\arcmin00.0\arcsec (J2000). This field lies within
one of the regions of the GAMA (Galaxy and Mass Assembly) survey \citep{gama} so optical spectra, along with
additional multiwavelength data, are available for the majority of the
low--redshift sources. 

The observations were taken in parallel--mode, which uses the
Photodetector Array Camera and Spectrometer \citep[PACS;][]{pacs} and
Spectral and Photometric Imaging REciever \citep[SPIRE;][]{spire}
instruments simultaneously; two orthogonal scans were used to mitigate 
the effects of $1/f$ noise.  The time--line data were reduced using {\tt HIPE} \citep{hipe}. SPIRE
250, 350, and 500\mic maps were produced using a na\"ive mapping
technique, after removing any instrumental temperature variations
(Pascale et al. \citeyear{spiremaps}), and incorporating the
appropriate flux calibration factors. Noise maps were generated by
using the two cross--scan measurements to estimate the noise per detector pass, and
then for each pixel the noise is scaled by the square root of the number of detector
passes. The SPIRE point spread function (PSF) for each band was determined from Gaussian fits to observations of Neptune, the primary calibrator for the instrument. 
Maps from the PACS 100 and 160\mic data were produced using the {\tt PhotProject} task within {\tt HIPE} 
(Ibar et al. \citeyear{pacsmaps}). A false colour combined image of
a part of the three SPIRE maps is shown in Figure
\ref{pre_backsub}. The measured beam full--width--half--maxima (FWHMs)
are approximately 9\arcsec, 13\arcsec, 18\arcsec, 25\arcsec and
35\arcsec for the 100, 160, 250, 350 and 500\mic bands respectively \citep{pacsmaps, spiremaps}. The map pixels are 2.5\arcsec, 5\arcsec, 5\arcsec,
10\arcsec and 10\arcsec in size for the same five bands. 

The noise levels measured by \citet{spiremaps} for the 250\mic and 500\mic SPIRE bands are in good
agreement with those predicted using the Herschel Space Observatory
Planning Tool (HSpot\footnote{{\tt HIPE} and HSpot are joint developments by
  the Herschel Science Ground Segment Consortium, consisting of ESA, the NASA Herschel Science Center, and the HIFI, PACS and 
SPIRE consortia}); for the 350\mic band they are considerably better. The corresponding PACS noise levels determined by \citet{pacsmaps} are currently
higher than predicted (26 mJy and 24 mJy, compared with 13.4 mJy and
18.9 mJy for 100\mic and 160\mic respectively), but this may
improve in future with better map--making techniques. The flux
calibration uncertainties are 15\% for the three SPIRE bands
\citep{spiremaps} and 10 and 20\% for the PACS 100\mic and 160\mic
bands respectively \citep{pacsmaps}.

\begin{figure*}
\centering
\subfloat[Source with a close companion]{\protect\label{bad_case} \includegraphics[scale=0.55]{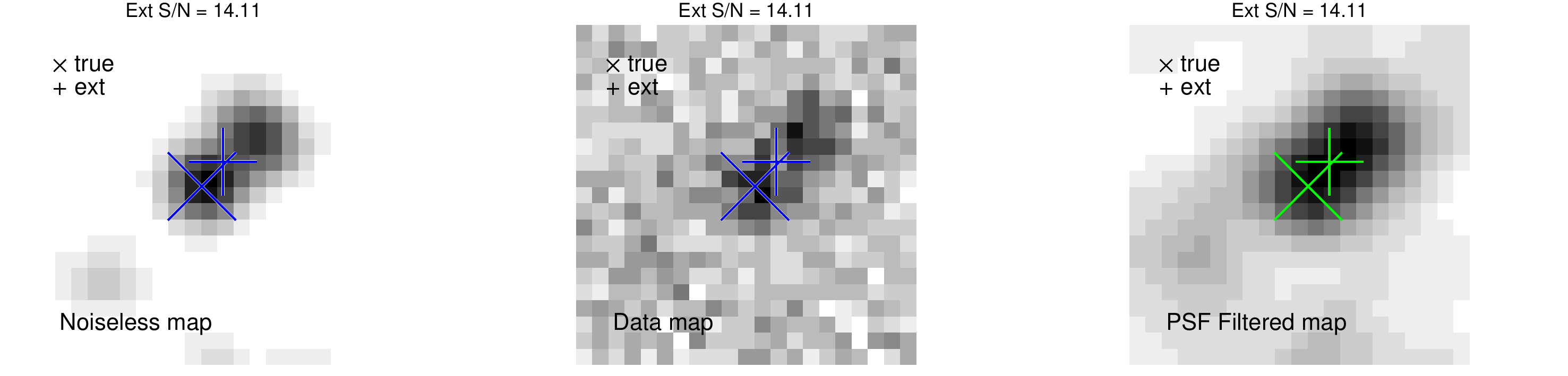}} \\
\subfloat[Isolated source]{\protect\label{good_case}\includegraphics[scale=0.55]{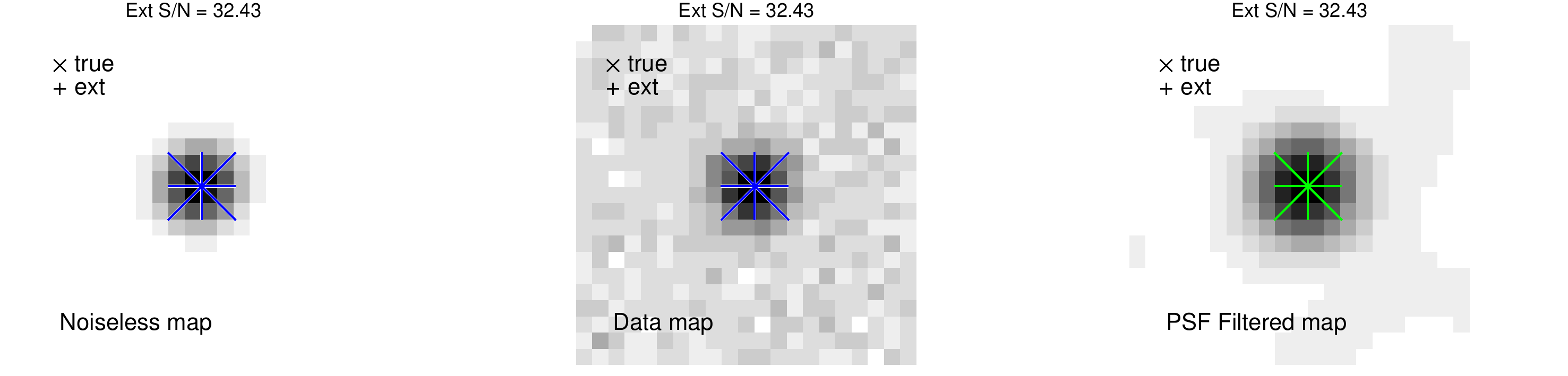}} \\
\caption{\protect\label{cutout_plots} The input (true) and extracted position
  for two point sources in the 250\mic simulated maps before the
  addition of Gaussian noise (noiseless), after the noise has been
  added (noisy), and after further convolving with the 250\mic point
  spread function (PSF) to create the final realistic sky
  (PSF--filtered) (see Section  \ref{sims} for full details), to
  illustrate how the position, and therefore flux density, of an extracted source found by MADX can be influenced by the presence of  a close companion.}
\end{figure*}

\begin{figure*}
\centering
\includegraphics[scale=0.7, clip, trim=0mm 4mm 0mm 2mm]{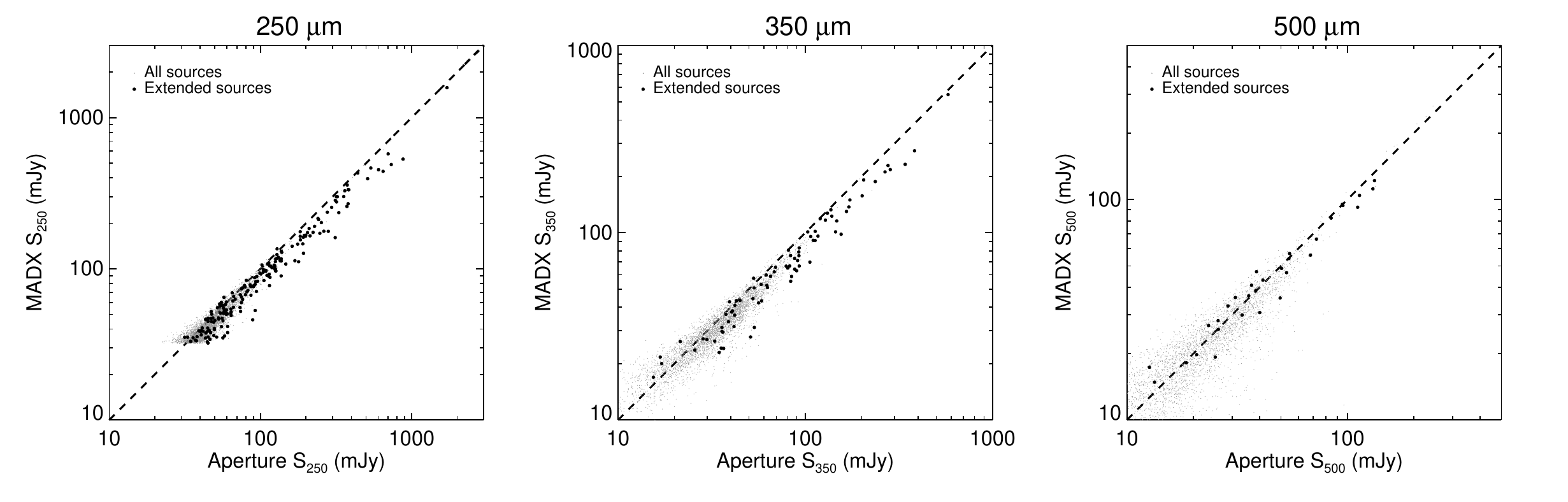}
\caption{\protect\label{ext_plots} A comparison between the MADX and
  aperture measured fluxes for the sources with a possible optical
  identification in the matched catalogue of \citet{Smith}. Source
  identified as extended are highlighted in bold.}
\end{figure*}

\section{Source extraction}
\label{extract}

The ultimate aim for the source identification of the H-ATLAS data is
to use a multiband method to perform extraction across the five
wavebands simultaneously, thus utilising all the available data as
well as easily obtaining complete flux density information for each
detected galaxy, without having to match catalogues between
bands. However, the short timescale for the reduction of these SDP
observations, combined with the higher than expected PACS noise
levels, means that this was only possible for the three SPIRE
bands. As a result, the source extraction for the PACS and SPIRE maps
is discussed separately in this Section.

The full H-ATLAS SDP catalogue described here will be available at
\verb1http://www.h-atlas.org/1.

\subsection{The SPIRE catalogue}
\label{spire_cat}

Sources are identified in the SPIRE 250, 350 and 500\mic maps using
the Multi--band Algorithm for source eXtraction \citep[MADX,][]{madx},
which is being developed for the H--ATLAS survey. Several methods for
generating the final SPIRE catalogue with MADX were investigated and these
are described below. 

The first step in the MADX source extraction is to subtract a local background, estimated from the peak
of the histogram of pixel values in $30\times 30$ pixel blocks (chosen
to allow the map to be easily divided up into independent sub--regions). This
corresponds to 2.5\arcmin $\times$ 2.5\arcmin for the 250\mic map, and
5\arcmin $\times$ 5\arcmin for the 350 and 500\mic maps. The
background (in mJy/beam) at each pixel was then estimated using a
bi-cubic interpolation between the coarse grid of backgrounds, and
subtracted from the data. 
Figure \ref{pre_backsub} illustrates the
reduction in background contamination (mainly arising from galactic
cirrus, which dominates over the confusion noise from unresolved sources) obtained using this method.

The background subtracted maps were then filtered by the estimated
PSF, including an inverse variance weighting, where the noise for each
map pixel was estimated from the noise map  \citep[matched filtering,
  e.g.][]{turin, serjeant}. 
The background removal has a negligible effect on the PSF because the histogram peak is insensitive to resolved
sources in the background aperture; this will be discussed further in \citet{madx}.
We also create a `filtered noise' map which represents the noise on a
pixel in the PSF filtered map. This is lower than the raw noise map
because the noise in the SPIRE pixels is uncorrelated, and so
filtering by the PSF reduces the noise by approximately the square
root of the number of pixels per beam.

The maps from the 350 and 500\mic bands are interpolated onto the 250\mic
pixels. Then all three maps are combined with weights set by the local
inverse variance, and the prior expectation of the spectral
energy distribution (SED) of the galaxies. 
We used two SED priors: a flat-spectrum prior (assumed to be flat in $f_{\nu}$), where equal weight is given to each band; and
also 250\mic weighting, where only the 250\mic band was included. 

Local, $>2.5\sigma$, peaks are identified in the combined PSF filtered
map as potential sources, and sorted in order of decreasing
significance level. A Gaussian is fitted to each peak in turn to provide an
estimate of the position at the sub--pixel level; this can be influenced by the presence of a
neighbouring source, as illustrated in Figure \ref{cutout_plots}, but
the effect is minimal. The flux in each band is then estimated using a bi--cubic interpolation to the position
given by the combined map. The scaled PSF is then subtracted from the
map before going on to the next source in the sequence. This ensures that
flux from the wings of bright sources does not contaminate nearby
fainter sources.  This sorting and PSF subtraction reduces the effect of confusion, but in future
releases we plan to implement multi-source fitting to blended sources.

To produce a catalogue of reliable sources, a source is only included
if it is detected at a significance of at least 5$\sigma$ in one of
the SPIRE bands. The total number of sources in the SPIRE catalogue is
6876.

For our current data we chose to use the 250\mic only prior for all our
catalogues, which means that sources are identified at 250\mic
only. At the depth of the filtered maps source confusion is a significant
problem, and the higher resolution of the 250\mic maps outweighed the
signal--to--noise gain from including the other bands (see Section
\ref{sim_creation} and Figure \ref{pos_err_pss_prior}). This may introduce a bias in the catalogue against red,
potentially high--redshift, sources that are bright at 500\mic, but
weak in the other bands. However, comparing catalogues made with both
the 250\mic and flat--spectrum priors showed that the number of missed sources is low: 2974
$>5\sigma$ 350\mic sources and 348 $>5\sigma$ 500\mic sources are
detected with the flat prior, compared with 2758 and 307 sources
detected using the 250\mic prior (i.e. 7\% and 12\% of sources are
missed at 350\mic and 500\mic respectively). It should also be noted that for a
high--redshift source to be missed it would need a 500\mic to 250\mic
flux ratio of $>2.7$ (i.e. it has to be $<2.5\sigma$ at 250\mic to be
excluded from the catalogue). Assuming typical SED templates (e.g. M82
and Arp220), this means that this should only occur for sources which lie at redshifts $>4.6$.
We aim to revisit this issue in future data--releases.

Since MADX uses a bicubic interpolation to estimate the peak flux in
the PSF filtered map, it partially avoids the peak suppression caused
by pixelating the time-line data, as discussed by Pascale et al.
Nevertheless the peak fluxes are systematically underestimated, and so
pixelization correction factors were calculated by pixelating the PSF
at a large number of random sub-pixel positions. The mean correction
factors were found to be 1.05, 1.11 and 1.04 in the 250, 350 and 500\mic
bands respectively, and they have been included in the released SDP catalogue.  

In calculating the $\sigma$ for each source, we use the filtered noise
map and add the confusion noise to this in quadrature. The average
1$\sigma$ instrumental noise values are 4.1, 4.0 and 5.7 mJy/beam
respectively, with 5\% uncertainty, in the 250, 350 and 500\mic bands, determined from the
filtered maps \citep{spiremaps}. We estimated the confusion noise from the difference
between the variance of the maps and the expected variance due to
instrumental noise (assuming that confusion is dominating the excess noise), and find that the 1$\sigma$ confusion noise is 5.3,
6.4 and 6.7 mJy/beam at 250, 350 and 500\mic, with an uncertainty of 7\%; these values are in good
agreement with those found by \citet{nguyen} using data from the
Herschel Multi--tiered Extragalactic Survey (HerMES). The resulting
average 5$\sigma$ limits are therefore 33.5, 37.7 and 44.0 mJy/beam.

\begin{figure}
\centering
\includegraphics[scale=0.4, clip, trim=0mm 0mm 0mm 0mm]{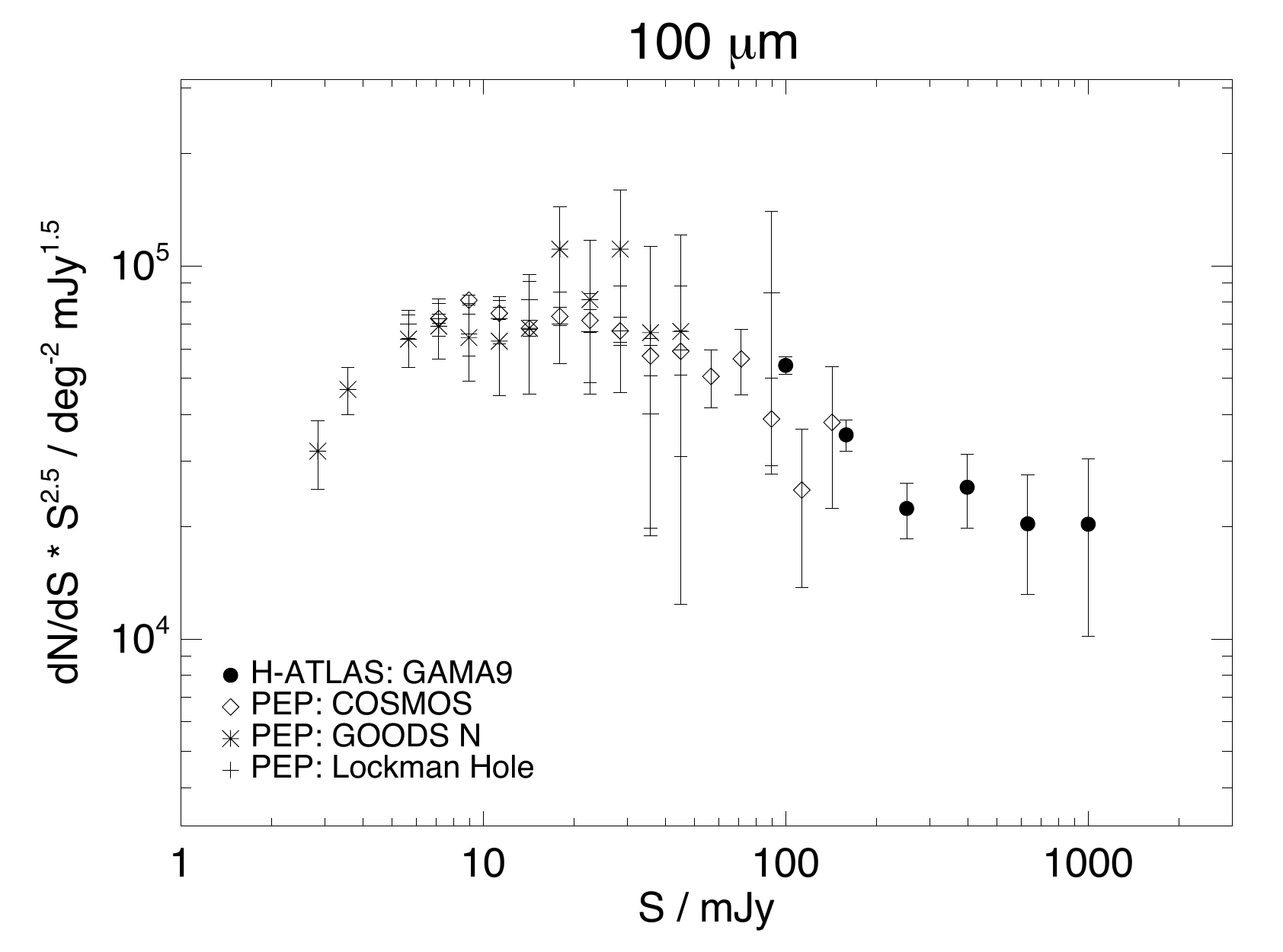}

\includegraphics[scale=0.4, clip, trim=0mm 4mm 0mm 0mm]{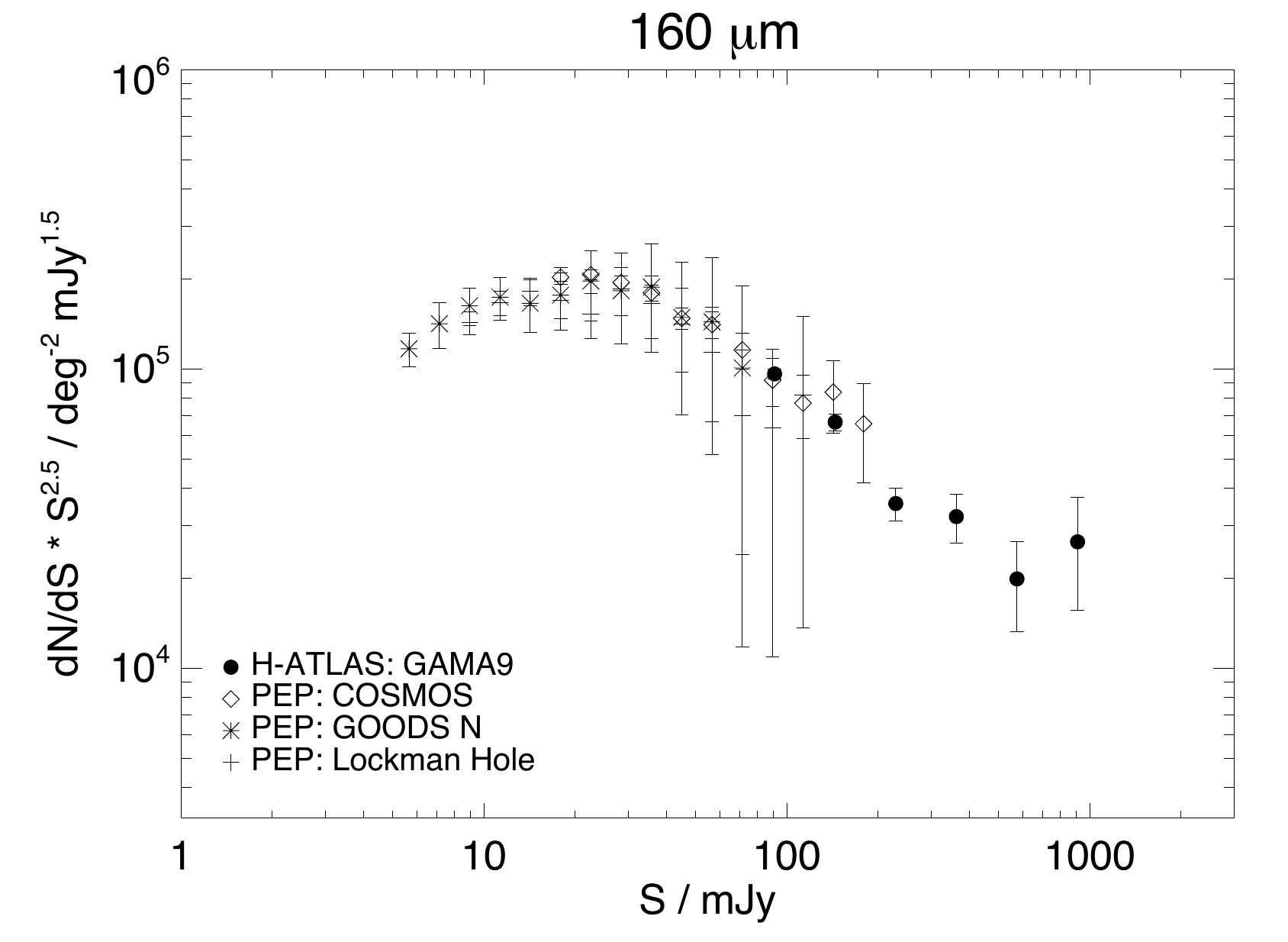}
\caption{\protect\label{pacs_cnts} The differential source counts from
  the PACS section of the SDP catalogue compared to the initial
  results from the three fields covered by the PEP survey
  \protect\citep{berta}. } 

\includegraphics[angle=180,scale=0.6, clip, trim=0mm 6mm 0mm 0mm]{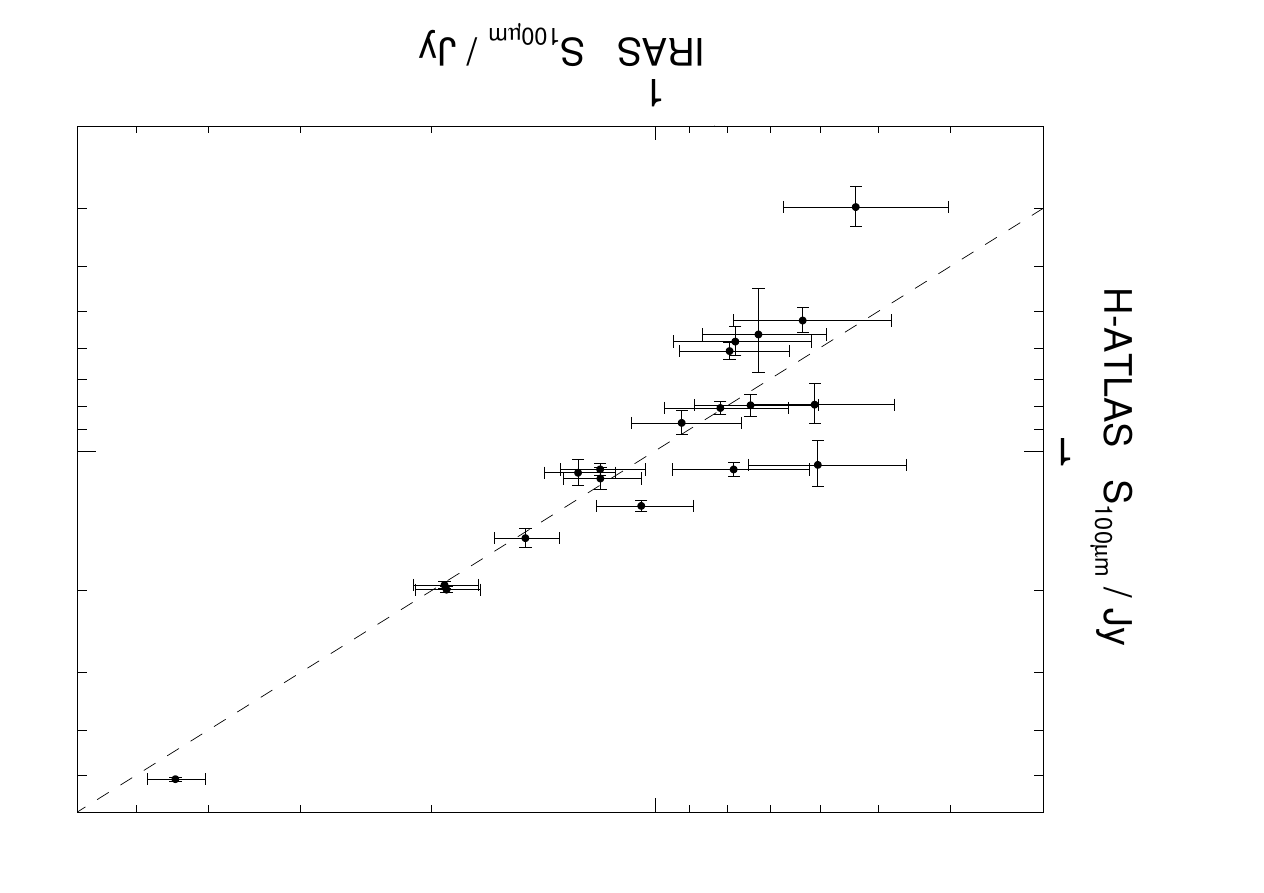}
\caption{\protect\label{iras_comp} A comparison between the 100\mic flux densities from PACS and IRAS}
\end{figure}

\subsubsection{Extended sources}
\label{ext_sour}

The flux density extracted by MADX will underestimate the true value for sources that are larger than the
SPIRE beams, which have FWHM of 18.1\arcsec, 24.8\arcsec and 35.2\arcsec for 250, 350 and 500\mic
respectively. This occurs because the peak value taken by MADX only
accurately represents the true flux density of a source if it is
point--like. These extended sources can be identified if they also 
have a reliable optical match and therefore a corresponding optical
size, $r_{\rm opt}$ (equivalent to the 25 mag arcsec$^{-2}$ isophote), in the SDSS or GAMA catalogues (see \citet{Smith} for
full details of the matching procedure and the determination of the
match reliability, $R_{j}$). The size of the aperture used is listed in the catalogue, and the most
appropriate flux density, either point source or aperture measurement
(when this is larger), is given for each source in the SPIRE `BEST flux'
columns.  It should be noted that, apart from two exceptions, this is necessary at 250 and 350\mic only, as the large 500\mic beam size means that the
flux discrepancy is negligible for that map. 

An `extended source', in a particular map, is defined here as one with $r_{\rm opt} > 0.5\times$FWHM, and to ensure only true matches are used, it must also have a
match--reliability, $R_{j}$, greater than 0.8. In total, the MADX `BEST' flux columns for 167 sources at 250\mic and 53
sources at 350\mic were updated with aperture photometry values. 

The aperture radius, $a_{r}$, in a particular band is set by summing the optical size in quadrature with the FWHM of that band:
\begin{equation}
\label{ap_rad}
a_{r} = \sqrt{{\rm FWHM}^{2} + r_{\rm opt}^{2}} . 
\end{equation}
The exceptions to this were the apertures used for sources H--ATLAS J091448.7-003533 (a
merger, where the given $a_{\rm r}$ is insufficient to include the
second component) and H--ATLAS J090402.9+005436, which visual
inspection showed was clearly extended. In these cases the aperture
sizes used are chosen to match the extent of the sub--mm emission, and fluxes are replaced in the 500\mic band as well. 

The apertures are placed on the MADX, Jy/beam, background subtracted
maps, at the catalogue position for each source; the measured values are converted to the correct flux scale
by dividing by the area of the beam derived by \citet{spiremaps} for each map (13.9, 6.6 or 14.2 pixels
for 250, 350 and 500\mic respectively). The corresponding
1$\sigma$ error is given by $\sqrt{v_{\rm ap}}$, where
$v_{\rm ap}$ is the sum of the variances within
apertures placed in the same positions on the relevant variance maps. Confusion
noise estimates were again added in quadrature to these uncertainties; these were scaled according to the area of each individual aperture. 

Figure \ref{ext_plots} compares the MADX and aperture measured fluxes
for all catalogue sources with a possible optical identification. It
shows that the majority of objects are point--like, for which the
agreement between the two sets of fluxes is good. The sources
identified as extended are highlighted in bold, and it is clear that
MADX underestimates these at 250 and 350\mic if they are brighter than $\sim$100 mJy.

\subsection{The PACS catalogue}

The higher noise levels in the PACS maps, along with the shape of the
source SEDs, mean that all the PACS extragalactic sources should be
clearly detected in the SPIRE catalogue. Sources in the PACS data are
therefore identified by placing circular apertures at the SPIRE
250\mic positions in the 100\mic and 160\mic maps, after correcting
the PACS astrometry to match that of the 250\mic map (using the
sources present in both the SPIRE and PACS maps). There are two steps to this
source detection process: first a `point source' measurement is obtained for all SPIRE
positions using apertures with radii of 10\arcsec (100\mic) or
15\arcsec (160\mic); next additional aperture fluxes are found for
positions where a PACS source would satisfy the extended source
criteria  discussed in Section \ref{ext_sour}. Aperture radii in this
case are calculated using Equation \ref{ap_rad}, assuming FWHM of
8.7\arcsec and 13.1\arcsec for 100 and 160\mic respectively. These
FWHM values are calculated using rough modelling of the Vesta asteroid
as the full PACS PSFs are asymmetric \citep[see][for a full
  discussion]{pacsmaps}. 

The aggressive filtering used for these maps means that the large
scale structure in the cirrus has already been removed, but some noise
stripes remain. These are removed globally at 160\mic by subtracting a
background determined within 10$\times$10 pixel blocks. However, at
100\mic this global approach was found to introduce negative holes
around bright sources so the background value is determined for each
source individually using a local annulus with a width of 0.5 times
the aperture radius. 

Unlike SPIRE, the PACS maps have units of Jy/pixel so no beam
conversion is needed. However, the fluxes are divided by 1.09 (100\mic) or 1.29 (160\mic) as recommended by the PACS
Instrument Control Centre\footnote{see the scan mode release note,
  PICCÂ­MEÂ­TNÂ­0.35}. These scaling factors are now incorporated into
the data--reduction pipeline and have been applied to the public
release of the PACS SDP maps, along with the astrometry correction
needed to match that of the SPIRE 250\mic map (this correction
is $\sim$1\arcsec in both PACS bands). The fluxes are also aperture corrected,
using a correction determined from observations of a bright point--like source. The
1$\sigma$ errors are found using apertures randomly placed in the
maps; note that these errors scale with aperture size. The low confusion noise compared to SPIRE, plus the fast scan speed 
used in these observations, means that the integration time used in 
H-ATLAS is insufficient to provide confusion limited images with PACS. Full details of these observations can be found in \citet{pacsmaps}. 

\begin{figure}
\centering
\subfloat[Extended source simulations]{\protect\label{ext_source_cnts} \includegraphics[scale=0.7, clip, trim=0mm 2mm 0mm 4mm]{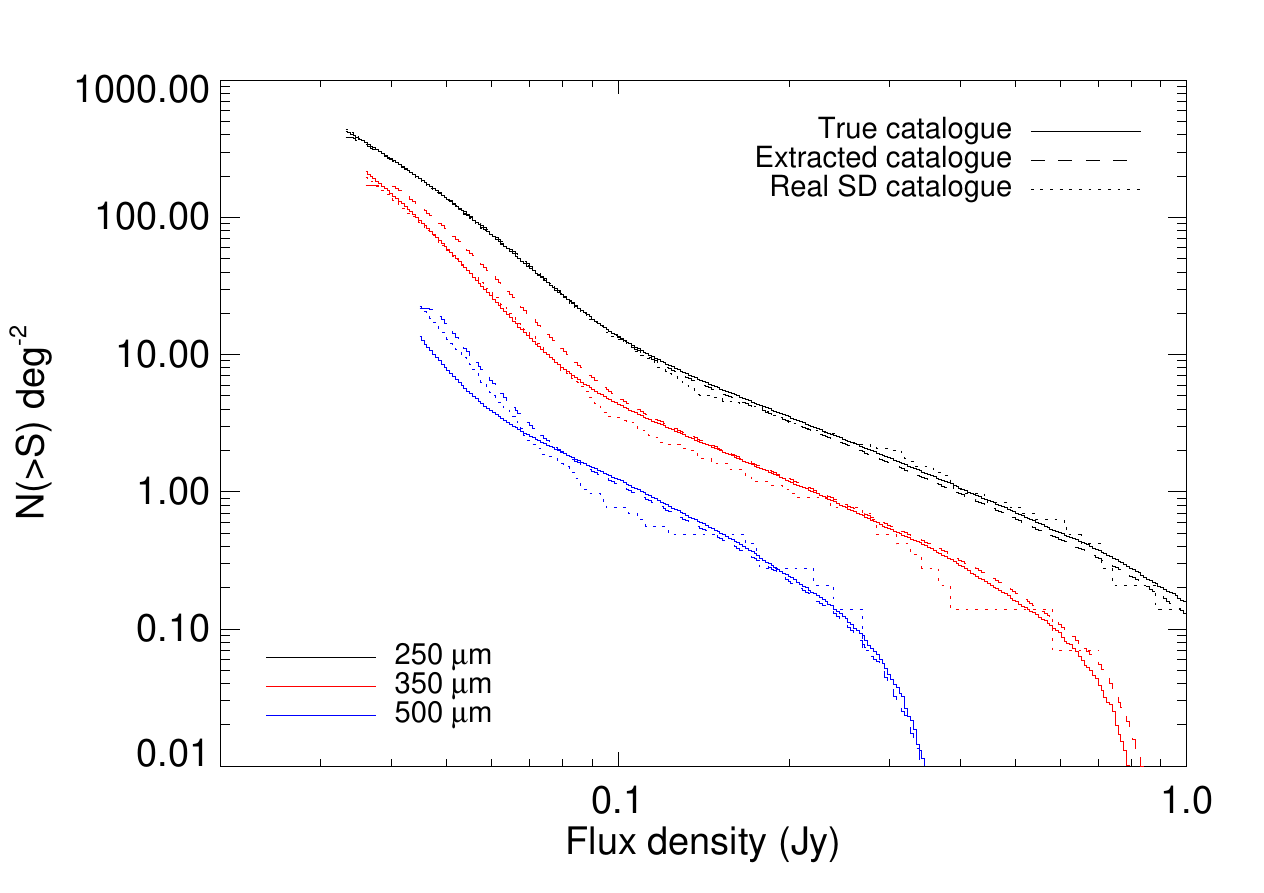}} \\
\subfloat[Point source only simulations]{\protect\label{pt_source_cnts} \includegraphics[scale=0.7, clip, trim=0mm 2mm 0mm 4mm]{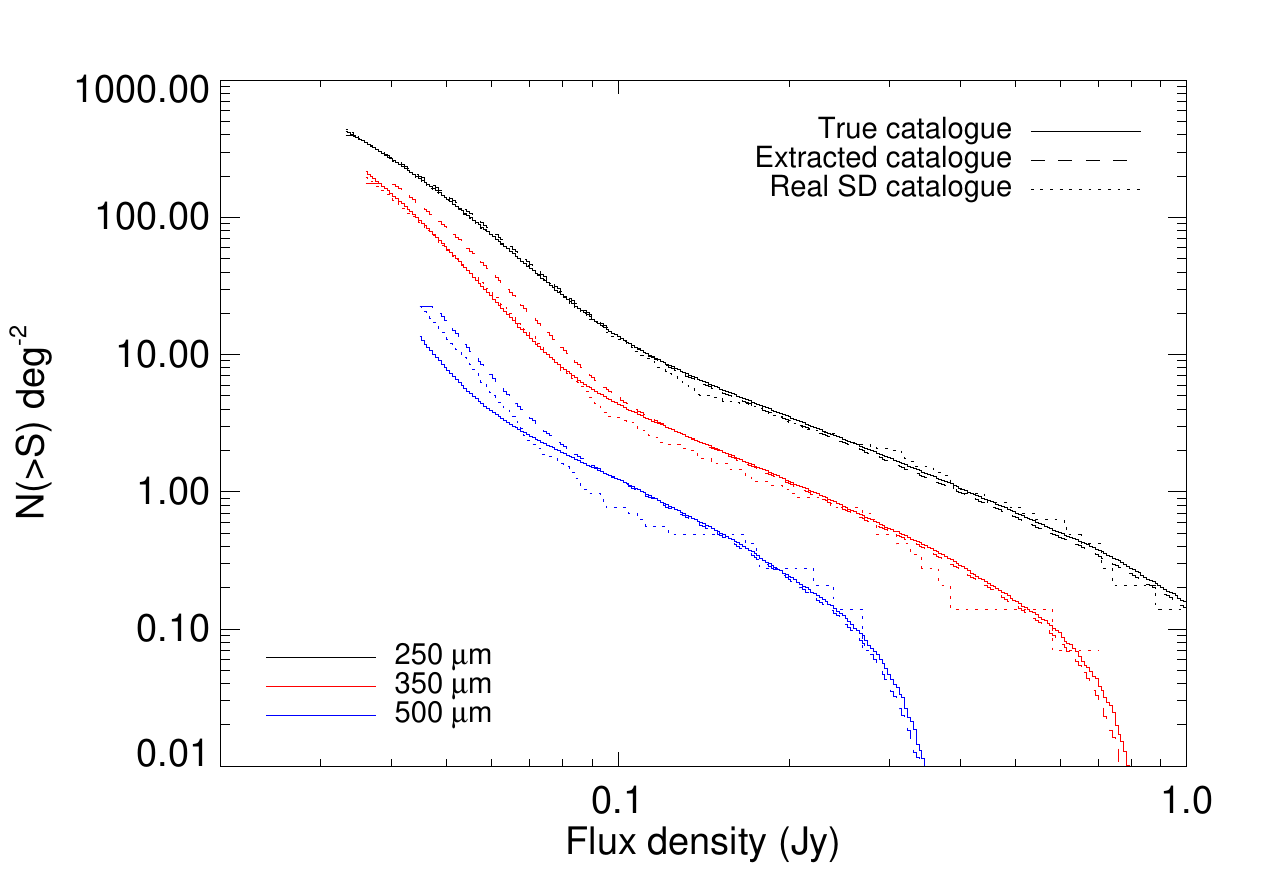}}
\caption{\protect\label{cnts_comp} The integrated source counts from
  the combined set of 500 input (true) and extracted simulated catalogues, along with those
  calculated using the SDP catalogue for both versions of the simulations. }
\end{figure}

\begin{figure}
\centering
\includegraphics[scale=0.6, clip, trim=0mm 0mm 0mm 0mm]{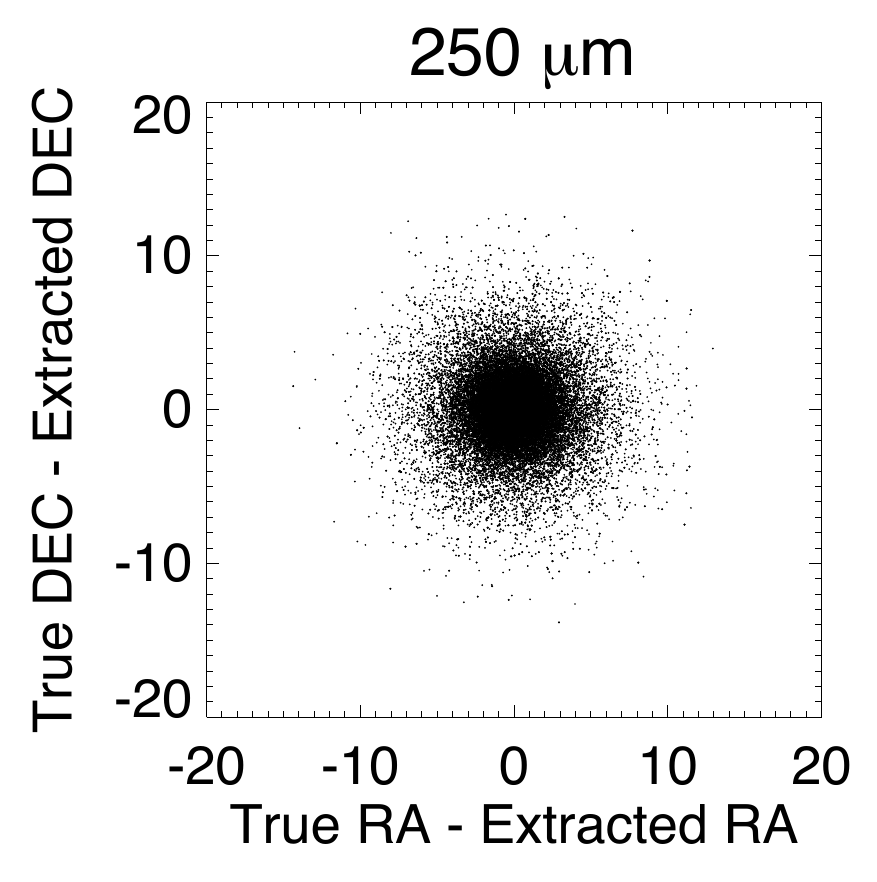}
\caption{\protect\label{xy_plots} The positional offsets between the
  matched sources in the simulated extracted and input (true) full 250\mic catalogues. The
  results for the two versions are very similar, so only the PSS
  points are shown here.}
\end{figure}

The most appropriate flux density measurements, either point or extended (where this is larger), are
given in the `BEST' PACS columns in the SDP catalogue, along with the
corresponding aperture radii, for sources with $S/N \geq 5$. As a
result 151 and 304 sources satisfy this condition at 100 and 160\mic
respectively. The 5$\sigma$ point source limits in the PACS
catalogue are 132 mJy and 121 mJy at 100 and 160\mic.  It should be
noted that the flux densities extracted from the PACS maps are only at 100\mic and 160\mic under the assumption of a  
constant energy spectrum, though the colour corrections for sources
with a different SED are small \citep{poglitsch}.

The PACS time-line data have been high-pass filtered by subtracting a
boxcar median over 3.4 arcmin (at 100\mic) and 2.5\arcmin at
160\mic \citep{pacsmaps}. The filtering will lead to the underestimation
of flux for sources extended on scales comparable to the filter length.
The exact flux loss for a particular source will depend on the size of
the source along the scan directions, and will also depend on whether
the peak surface brightness is above the 4$\sigma$ threshold used in
the second level filtering. A simple simulation of a circular exponential disc shows that the
filtering removes $\sim 50\%$ of the source flux if the diameter of the
disk is equal to the filter length. If the diameter is half of the
filter length, then only 5\% of the flux is removed. This suggests that
sources with a diameter less than 1\arcmin should by relatively
unaffected by the filtering. Flux measurements for sources larger than
this should be treated with caution.

Figure \ref{pacs_cnts} compares the differential source counts
calculated from the PACS SDP catalogues to those determined from the
initial data of the complementary PACS Evolutionary Probe (PEP) survey
\citep{berta}, which is deeper than H--ATLAS but covers a smaller
area. The good agreement between the two sets of counts supports the initial assumption that all bright PACS sources should
already be present in the SPIRE catalogue. However, there are
insufficient sources in the SDP data to properly constrain the bright
number counts tail. A full analysis of the PACS counts will be presented in \citep{ibar2}. 

\begin{figure*}
\centering
\subfloat[Extended source simulations]{\protect\label{pos_err_ess} \includegraphics[scale=0.5, clip, trim=3mm 5mm 0mm 0mm]{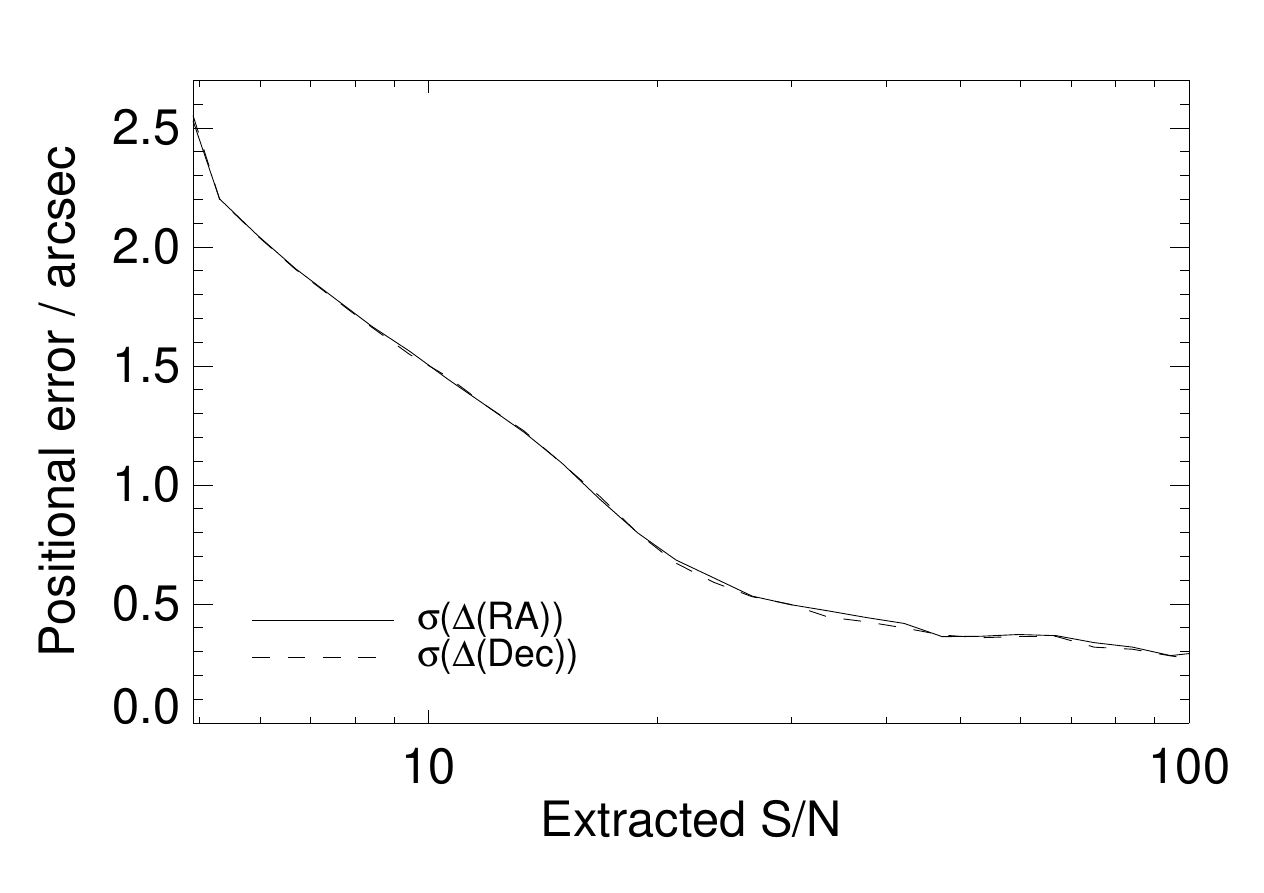}}% \\
\subfloat[Point source simulations]{\protect\label{pos_err_pss} \includegraphics[scale=0.5, clip, trim=3mm 5mm 0mm 0mm]{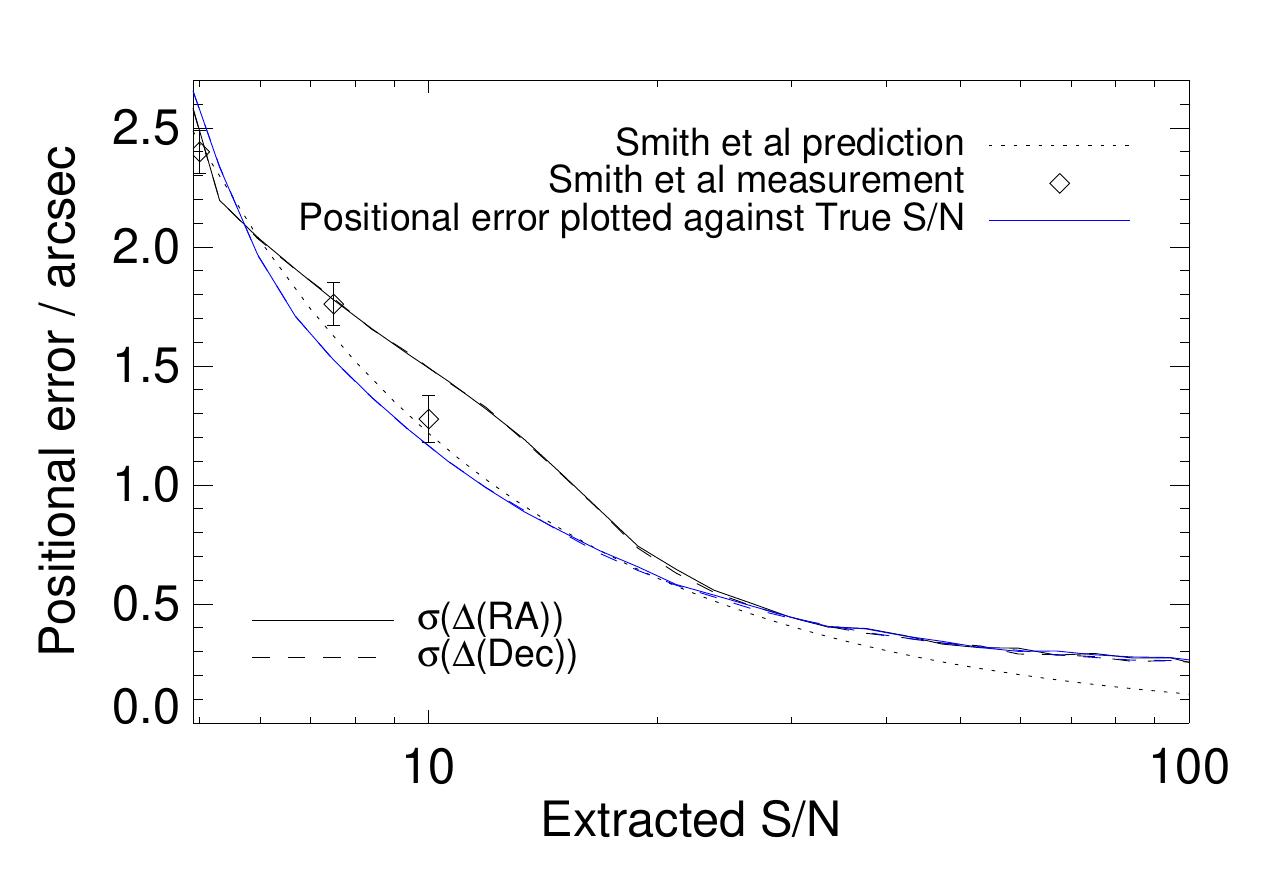}} 
\subfloat[Point source simulations comparing different priors]{\protect\label{pos_err_pss_prior} \includegraphics[scale=0.5, clip, trim=3mm 5mm 0mm 0mm]{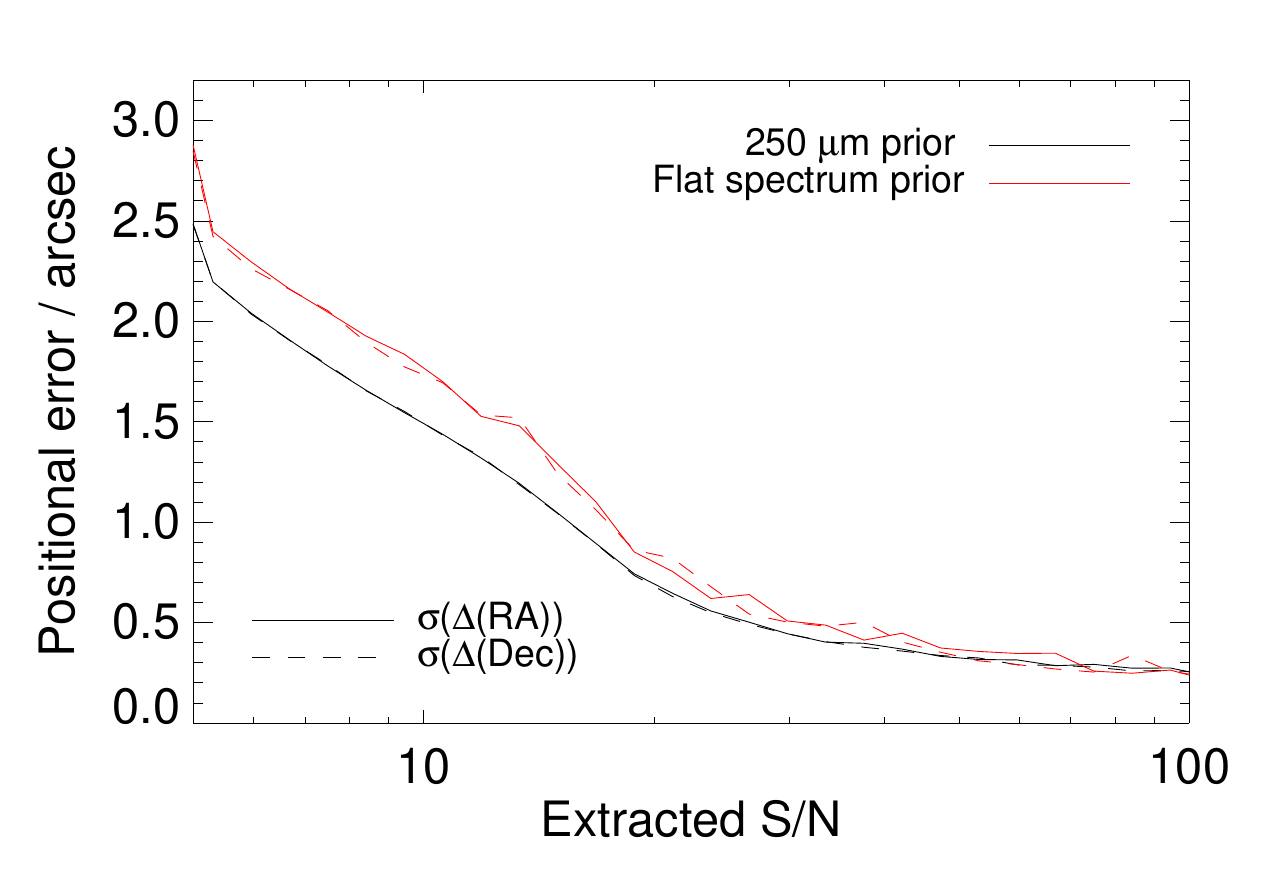}}
\caption{\protect\label{pos_err} The positional errors 
  for the two different versions of the simulations, alongside a
  comparison of the two different source extraction position priors
  as previously discussed in Section
  \ref{spire_cat}. Also shown in \ref{pos_err_pss} are the positional
  errors plotted against the S/N in the input (true) catalogue, along with
  those determined by \citet{Smith} for the SDP data at 5, 7.5 and
  10$\sigma$.}   
\end{figure*}

For the sources detected in the PACS 100\mic map an additional comparison can be made to this wavelength in
the Imperial IRAS--FSC Redshift Catalogue of \citet{iras}, which combines the original IRAS Faint Source Catalogue flux density values
with improved optical and radio identifications and redshifts. There are 34 IRAS sources within the PACS region of the
H-ATLAS SDP field; 19 of these have a reliable IRAS flux measurement and
these are in good agreement with the SDP catalogue, with a mean offset
consistent with zero, as shown in Figure \ref{iras_comp}.

\begin{figure*}
\centering
\subfloat[Extended source simulations]{\protect\label{ext_true_plots_ess} \includegraphics[scale=0.65, clip, trim=10mm 0mm 0mm 0mm]{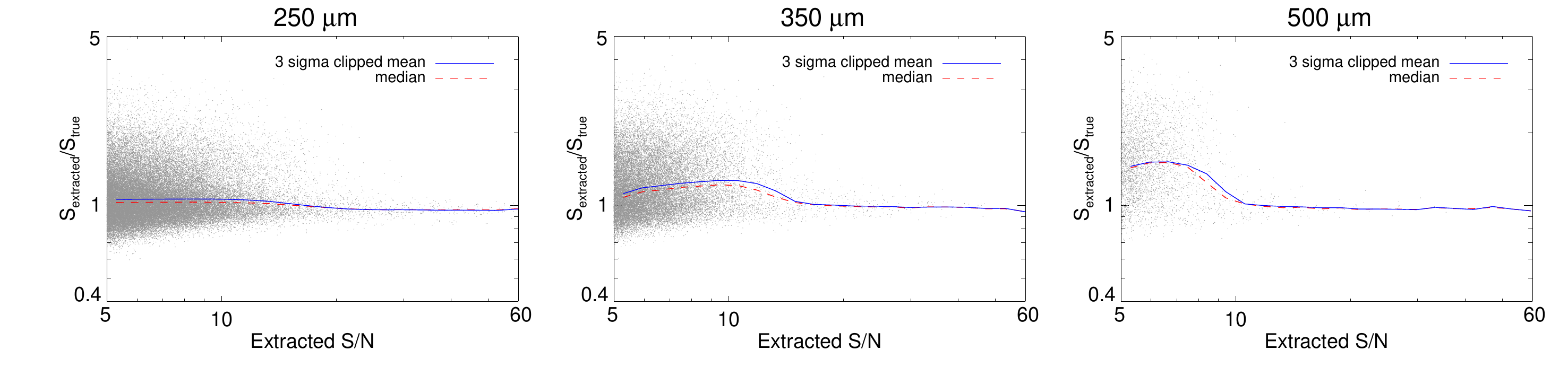}} \\
\subfloat[Point source simulations]{\protect\label{ext_true_plots_pss} \includegraphics[scale=0.65, clip, trim=10mm 0mm 0mm 0mm]{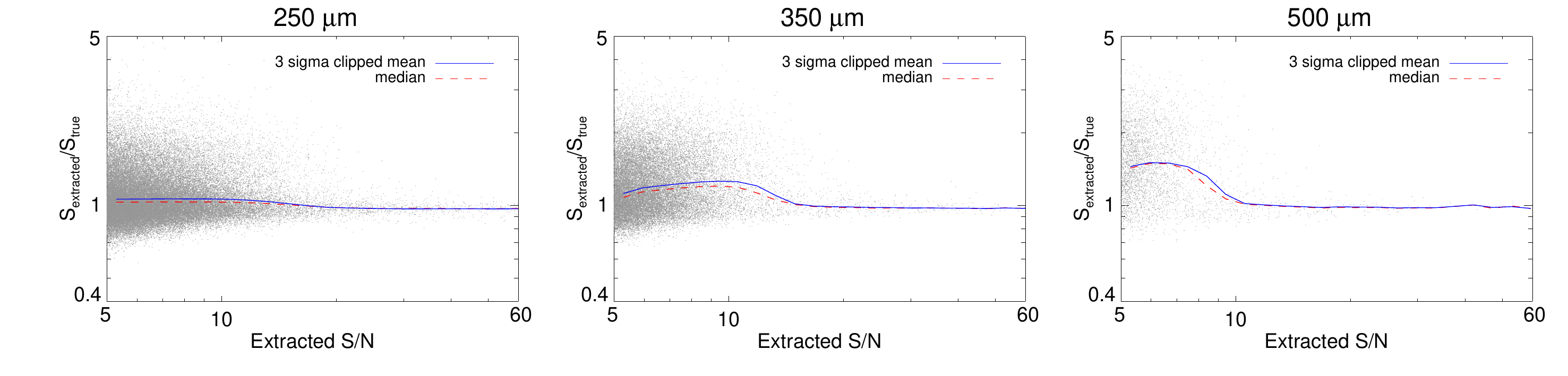}}
\caption{\protect\label{ext_true_plots} The ratio of flux densities
  for the matched sources in the simulated input (true) and extracted catalogues as a
  function of extracted signal to noise (S/N) for the three
  bands from the ESS and PSS maps. Also shown are the median and 3$\sigma$ clipped mean values,
  calculated in bins of 0.05 in $\log(S/N)$.}

\includegraphics[scale=0.65, clip, trim=10mm 0mm 0mm 0mm]{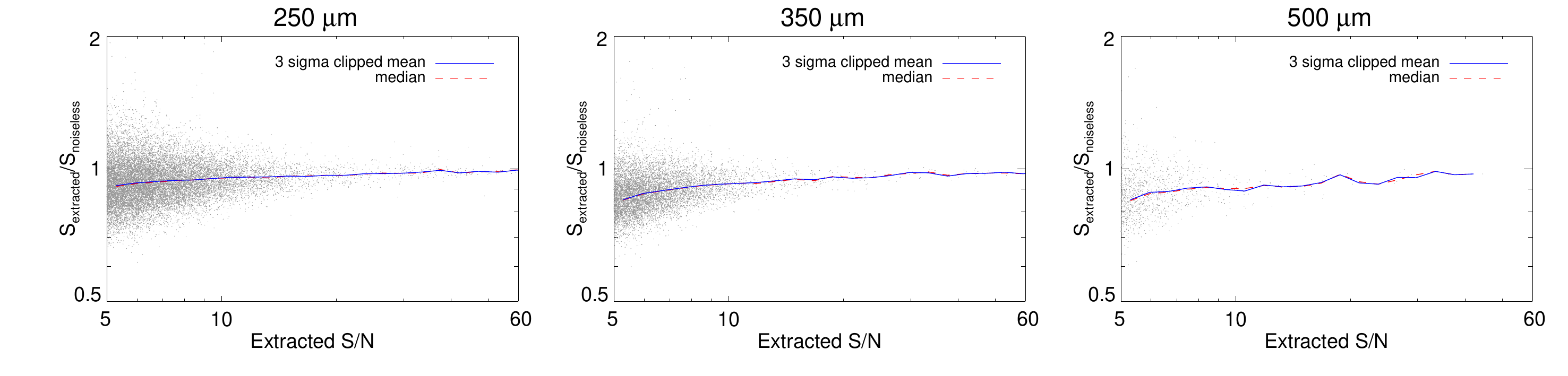}
\caption{\protect\label{ext_true_noiseless_plots} The ratio of flux
  densities for the matched sources in the noiseless MADX ($S_{\rm noiseless}$) and extracted
  catalogues as a function of extracted signal to noise for the
  three bands (including point sources only). Also shown are
  the median and 3$\sigma$ clipped mean values, calculated in bins of
  0.05 in $\log(S/N)$.} 

\includegraphics[scale=0.65, clip, trim=10mm 0mm 0mm 0mm]{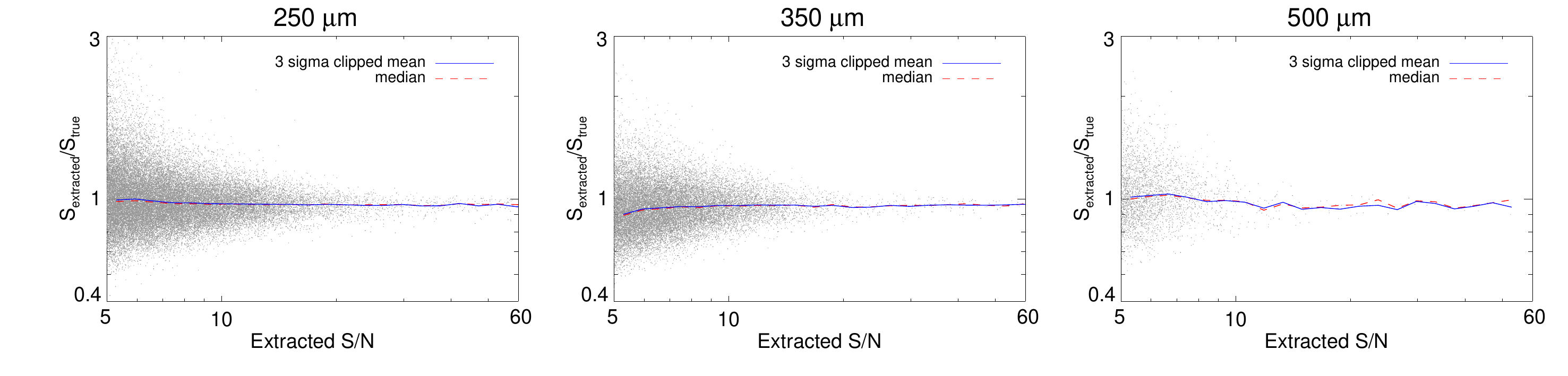}
\caption{\protect\label{ext_true_grid_plots} The ratio of flux
  densities for the matched sources in the simulated input (true) and extracted
  catalogues as a function of extracted signal to noise for the
  three bands, using the gridded position simulations (including point sources only). Also shown are
  the median and 3$\sigma$ clipped mean values, calculated in bins of 0.05 in $\log(S/N)$. Note that confusion noise is not included in these simulations.}

\end{figure*}

\begin{figure*}
\centering
\includegraphics[scale=0.6, clip, trim=6mm 0mm 0mm 0mm]{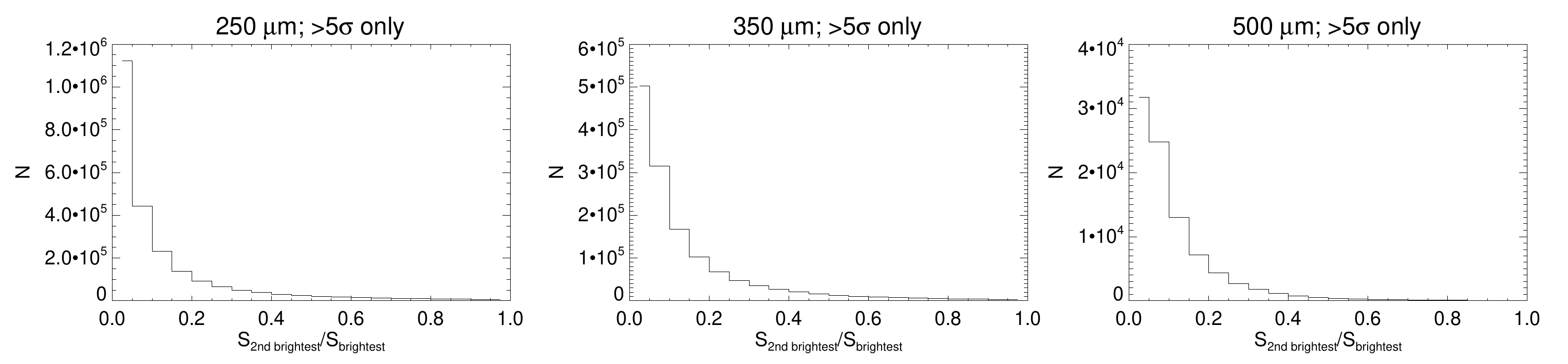}
\caption{\protect\label{beam_hists} The PSF--weighted ratio of the brightest to
  second brightest input (true) source contributing to the extracted source,
  within the beam in each band, for $>5\sigma$ sources in the extracted catalogue. }

\includegraphics[scale=0.6, clip, trim=8mm 0mm 0mm 0mm]{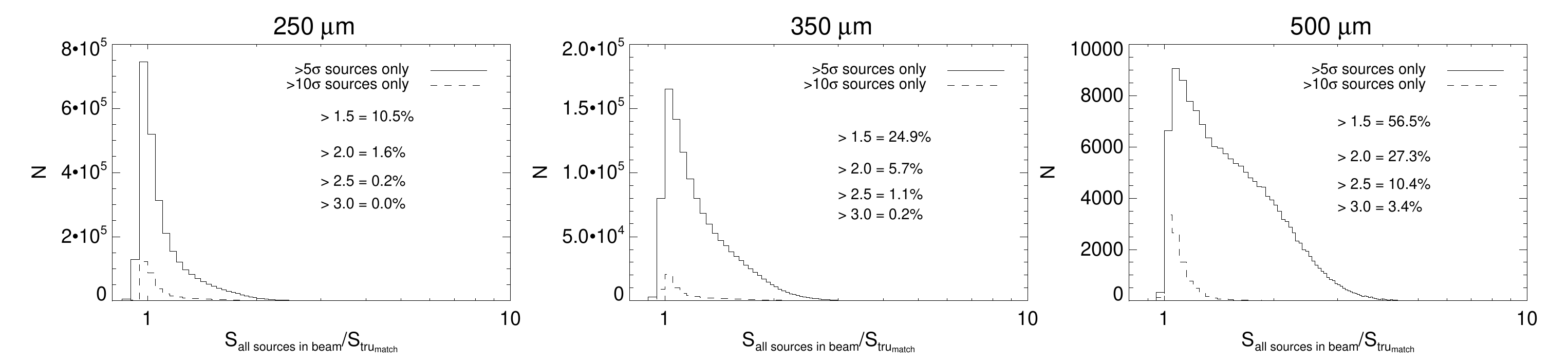}
\caption{\protect\label{beam_hists_total} The PSF--weighted, background--subtracted, ratio of
  the sum of simulated input (true) sources within a beam to the flux density of the
  matched true source for $>5\sigma$ (solid line), and $>10\sigma$
  (dashed line) sources in the extracted catalogue. The labels on the
  Figures give the percentage of sources with ratios greater than some
  particular value. The small proportion of sources where the ratio falls
  below 1 are due to the PSF--weighting. }  
\end{figure*}

\section{Assessing the catalogue reliability}
\label{sims}

\subsection{Simulation creation}
\label{sim_creation}

It is not enough to identify sources in the H-ATLAS SDP maps; the
robustness of the catalogue must also be determined. This is done
using realistic simulations of the observations, with the same noise
properties as the processed maps, and a realistic cirrus background,
based on IRAS measurements \citep{iras_ref}. However, only the three SPIRE bands are considered in this initial
analysis, as the PACS SDP catalogue is currently treated as an
extension to the SPIRE data

The simulated maps are randomly populated with sources generated using
the models of \citet{negrello}, which predict the number counts of
both the spheroidal and protospheroidal galaxy populations separately;
for the simulations, these predictions are combined together to give
the expected total counts, and hence the corresponding set of source
flux densities, for each band. Although \citet{wtheta} detected, in SDP data, strong
clustering for $350\,\mu$m and $500\,\mu$m--selected samples,
fluctuations due to faint sources at the SPIRE resolution are Poisson dominated,
especially at $250\,\mu$m \citep[e.g.][]{negrello04, viero09}. This
suggests that, for the present purposes, using unclustered random positions
is a sufficiently good approximation. The flux densities of all the sources in the models are reduced by 26\% at 250\mic and
15\% at 350\mic to improve the agreement with the observed (i.e. uncorrected) source counts
in the SDP catalogue \citep{clements}; the results of this alteration
are shown in Figure \ref{cnts_comp}. The final flux density ranges are 0.11 mJy -- 1.65 Jy
at 250\mic, 0.24 mJy -- 0.83 Jy at 350\mic and 0.45 mJy -- 0.59 Jy at
500\mic for the simulated sources; this ensures that the simulated
maps contain a realistic background of faint sources which can
contribute to the confusion noise.  

The simulations are constructed by first adding the flux of each
source in each band to the relevant position in a 1 arcsecond grid.  Two versions
of the simulations are created. In the first the simulated sources are
all one pixel in size (point--source--simulations: PSS), whereas in
the second the sources are assigned a scale--length based on their
catalogue redshift (extended--source--simulations: ESS). The
scale--length is constant in physical units, and then converted to
an angular scale using standard cosmology. The ESS will obviously be a better representation of the real data, but, as
Section \ref{ext_sour} shows, MADX underestimates the flux densities of objects with sizes larger than the FWHM, 
so the PSS simulations provide a useful comparison. It should be noted that the flux densities and
positions of the input sources will be the same in both cases.  The
next step is to convolve the 1 arcsecond map by the appropriate {\it Herschel} PSF, also
sampled on a 1 arcsecond grid, to give a map of flux per beam covering
the full area of the SDP data. Then, the 1 arcsecond pixels are block
averaged to give 5 arcsecond pixels for the 250\mic maps, and 10
arcsecond pixels for the 350 and 500\mic maps. 

A background representing emission from Galactic cirrus is then added
to the each map. The background value is estimated from the \citet{iras_ref} map of 100\mic dust
emission and temperature by assuming a modified black-body spectrum
with $\beta = 2.0$, and scaling to the appropriate wavelength. The resolution of this IRAS map is lower than that in the SDP data,
which means that small scale structure in the cirrus is not present in
the simulations. Since the cirrus is highly structured, it is
non--trivial to generate realistic structure on smaller scales, so as
a simple approximation, the low resolution maps were used, though it
should be noted that the true cirrus background will include more
small scale features. It is clear that the real cirrus structure in
the SDP data is highly non--Gaussian, so simply extrapolating the
power spectrum to smaller scales does not significantly improve the
model background. 

Finally instrumental noise is added to each pixel as a Gaussian
deviate, scaled using the real coverage maps so that the local rms is
the same as in the real data.

\begin{figure}
\centering
\subfloat[Extended source simulations]{\protect\label{flux_err_ess} \includegraphics[scale=0.6, clip, trim=0mm 4mm 0mm 4mm]{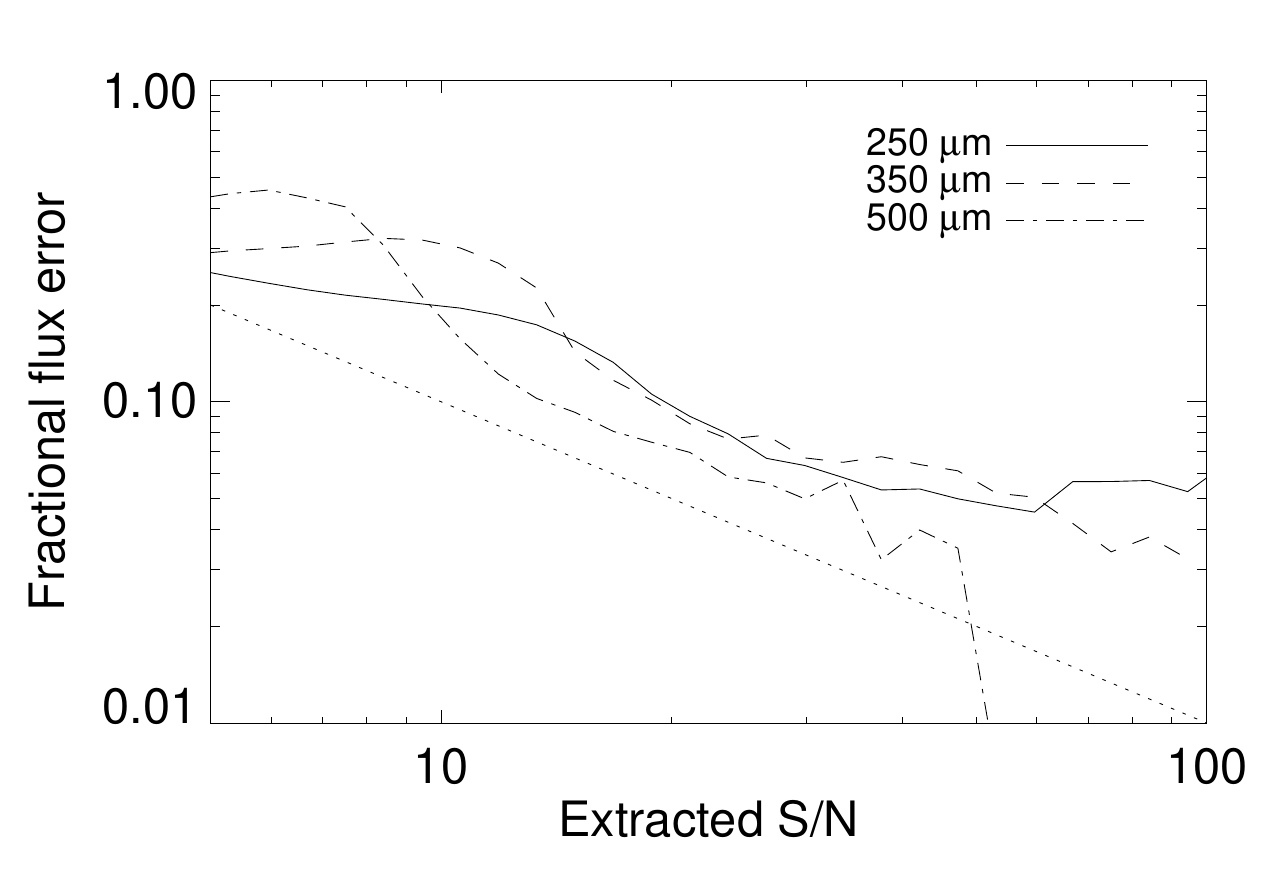}} \\
\subfloat[Point source simulations]{\protect\label{flux_err_pss} \includegraphics[scale=0.6, clip, trim=0mm 4mm 0mm 4mm]{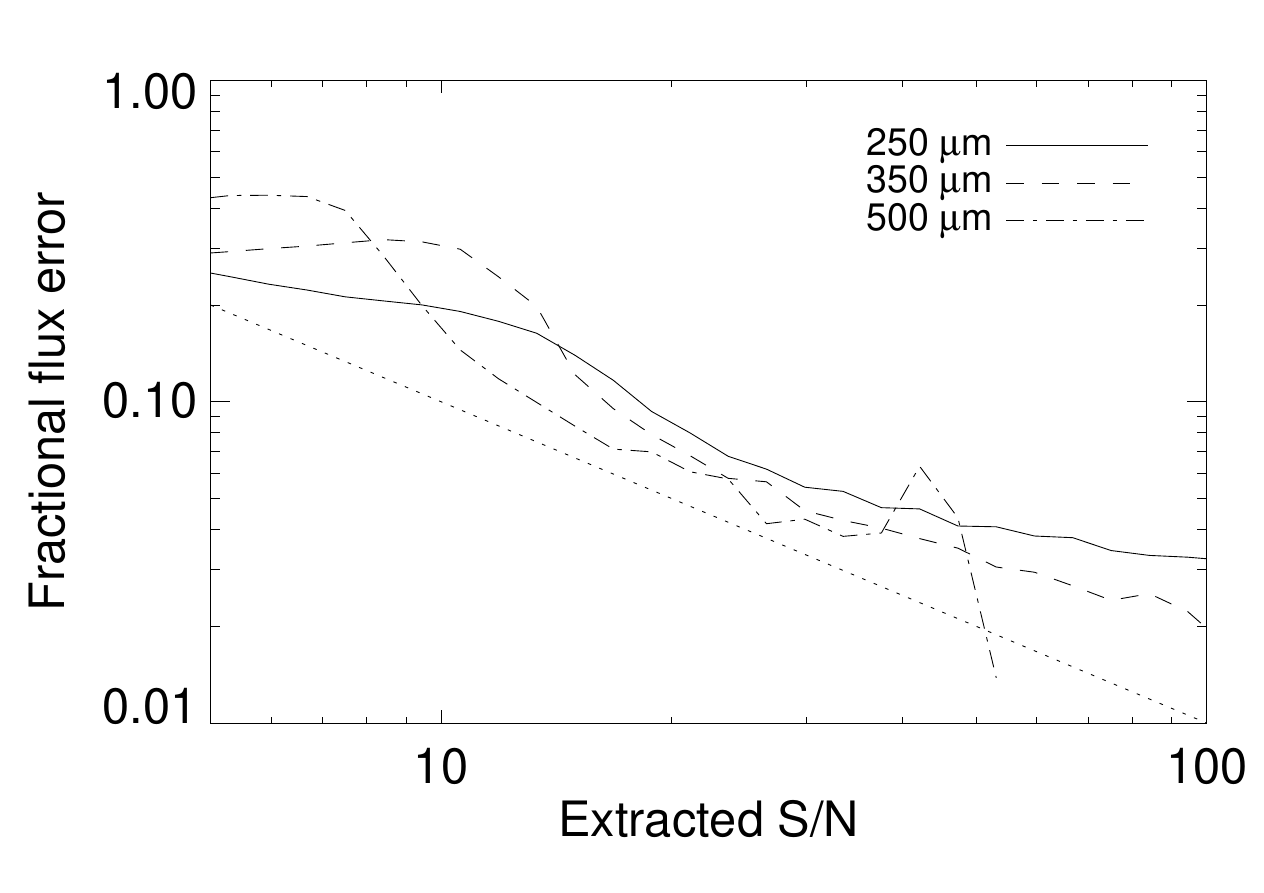}}\\
\subfloat[Point source simulations]{\protect\label{flux_err2_pss} \includegraphics[scale=0.6, clip, trim=0mm 4mm 0mm 4mm]{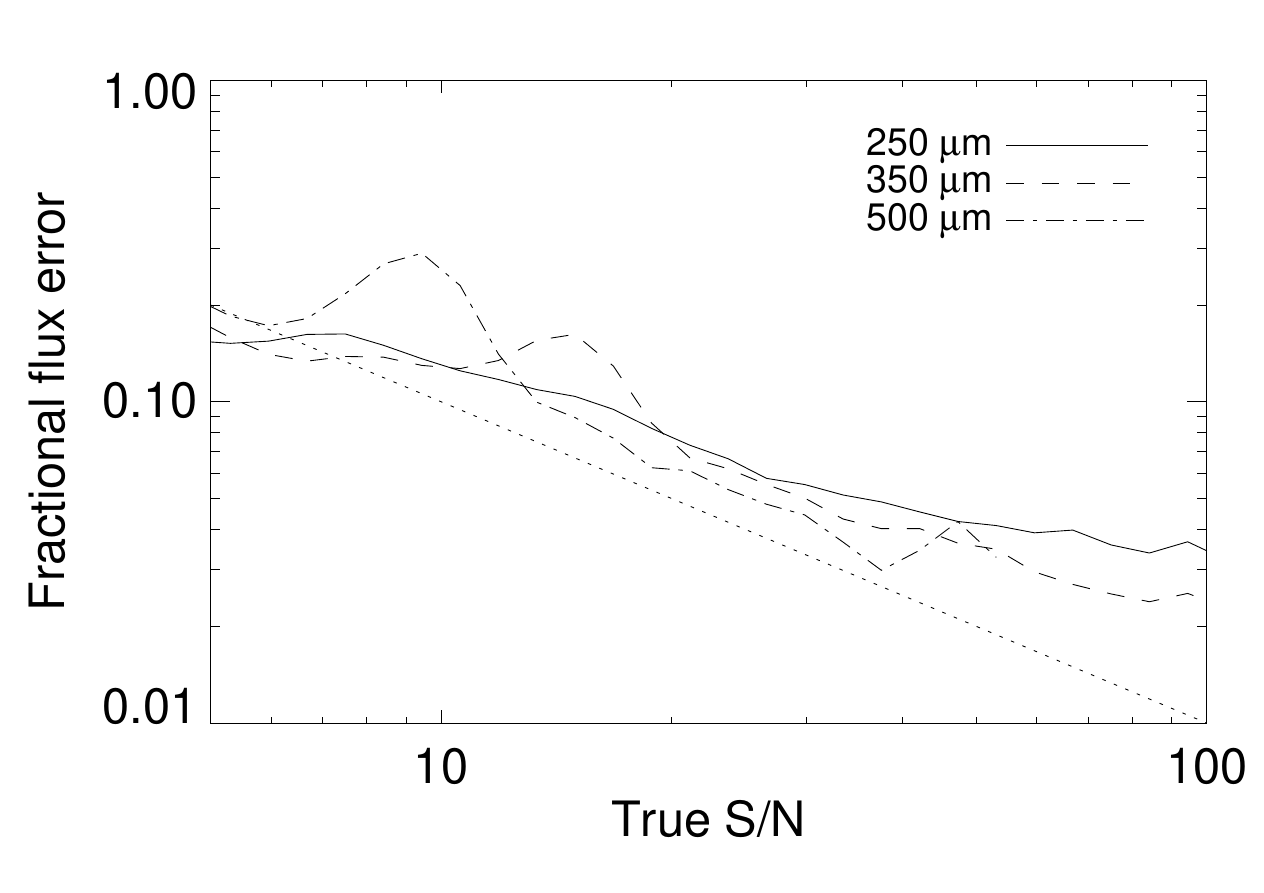}}
\caption{\protect\label{flux_err} The fractional flux density error
  for the corrected extracted catalogues, ignoring any sources that
  fall outside the 99.73rd percentiles. The dotted lines indicate the
  expected behaviour. 
}
\end{figure}

Sources are then extracted with MADX  from both versions of the
simulations, following the procedure described in Section
\ref{spire_cat}. For the ESS maps, the flux densities in the three bands are again
replaced with aperture--measured values for the extended sources. The
`optical sizes' (needed to determine $a_{r}$ using Equation
\ref{ap_rad}) in this case are taken as three times the scale--size
taken from the input catalogues; this corresponds to a {\it B}--band
isophotal limit of $\sim$25 mag arcsec$^{-2}$ \citep{zhong}. Finally, the MADX catalogue is cut to 
only include sources which are detected at the 5$\sigma$ level in any
of the available bands. This process is repeated 500 times, each time using a different realisation of
the input model counts, to ensure sufficient numbers of bright sources
are present at the longer wavelengths. The average number of extracted
sources which are also $>$5$\sigma$ in any band is 5881 and 5772 for
PSS and ESS respectively, which is lower than the 6876 sources present in the real SDP data; as Figure
\ref{cnts_comp} illustrates, this is because the simulated source
counts do not exactly reproduce the real SDP ones. Additionally, more
sources are found for the PSS version because of the flux
underestimation of extended sources which means that the faintest
objects fall below the catalogue cut. In the remainder of this
discussion, these MADX catalogues will be referred to as the `extracted catalogues', and the simulated input
source lists as the `simulated input catalogues'.    

For each of the three bands in turn, starting with the brightest,
sources in the extracted catalogue are matched to the simulated input source that
makes the largest contribution, determined by weighting with the
filtered beam, at that extracted position. A match radius of 3 pixels
(approximately equal to the FWHM in each band) is also imposed to
ensure that a match is not made to an unfeasibly distant source. Since
the typical positional error for a $>5\sigma$ 250\mic source is 2.5\arcsec or less, this match radius
will ensure that almost no real matches are rejected, whilst the weighting will avoid spurious matches. Once
matched, a simulated input source is removed from consideration to avoid double--matches. Considering
each band separately will allow an extracted source to have three
different simulated input counterparts, depending on where the majority of its
flux density comes from at 250, 350 and 500\mic. This ensures that the
effects of source blending in the data can be properly investigated,
though it should be noted that the results are very similar if the
counterparts are found at the highest resolution, shortest wavelength
only. Full simulated input, extracted and matched catalogues for each band are
then made by combining the results from the 500 individual sets of simulations together. 

The positional offsets and corresponding errors are shown in Figures \ref{xy_plots} and
\ref{pos_err}. They demonstrate that there is no significant offset
between the extracted and matched catalogues. The positional errors
for 5$\sigma$ sources are $\sim$2.4\arcsec at 250\mic in both
versions, which agrees with the value of $2.40 \pm 0.11$\arcsec found
for the real SDP data by \citet{Smith}. The errors also approximately scale as $1/(S/N)$ in the 250\mic band,
as predicted by e.g. \citet{ivison}. However, at low S/N there is an
enhancement over the predicted values, as illustrated for the PSS
results in Figure \ref{pos_err_pss}. This is a result of Eddington
bias causing more faint sources errors to scatter up than vice--versa; if the positional errors
are plotted against the S/N in the simulated input catalogue, which does not
suffer from this effect, then they are in better agreement with the
prediction.  

Figure \ref{pos_err_pss_prior} also illustrates the improvement in
positional errors that arises from selecting sources at 250\mic only
in MADX, instead of giving equal weight to all bands (flat--spectrum prior), as previously
discussed in Section \ref{spire_cat}. Greater positional accuracy
significantly enhances the efficacy of the cross--identification to
optical sources using the Likelihood Ratio method \citep{Smith}. This
is why the better positions are deemed to outweigh the slight chance
of missing red objects when using the 250\mic prior.

\subsection{Catalogue correction factors}

Inspection of Figure \ref{cnts_comp} shows a clear discrepancy between
the extracted and simulated input integral counts at faint 500\mic 
flux densities; this occurs due to a combination of two factors. The
first,  flux--boosting, is a preferential enhancement of faint source
flux densities due to positive noise peaks, that arises due to the
steepness of the faint end \citep[i.e. $S_{\rm 500\mu m} \lesssim 40$ mJy;][]{clements}  of the source counts. The second is a
result of blending, where several  simulated input sources (which may be too faint to be included individually) are detected as one
source in the extracted catalogue.

These effects can be quantified by direct comparison of the simulated input and extracted flux densities, shown in Figure
\ref{ext_true_plots}, as a function of signal--to--noise in the
extracted catalogue for both the ESS and PSS versions. Flux correction factors are derived from the
3$\sigma$ clipped mean of these data; these are given in Table
\ref{corr_table}. Applying these factors to each extracted source
gives a statistically `flux--corrected' catalogue.  It should be noted however, that
the discussion of correction factors in this Section is restricted to
sources detected at a 5$\sigma$ or greater level only. 

An alternative approach to determining the catalogue correction
factors is to use a `noiseless' catalogue, created by running MADX on
the simulated maps before the addition of noise, as the comparison. As Figure \ref{ext_true_noiseless_plots} shows, this does not accurately
represent the level of flux--enhancement in the data, because, the
noiseless catalogue is also affected by source blending. 
Additionally, at low S/N the noiseless--input flux densities are generally brighter
than the extracted ones, suggesting that MADX underestimates the
background subtraction in the absence of noise.  

The relative contributions from the flux--boosting and source blending
can be investigated with a new set of simulated, point--source only,
maps, in which the sources are placed on a regular spaced grid, with a
70\arcsec separation between points, to ensure no sources overlap. The source density is also
lowered in these maps (imposed by excluding any source in the
simulated input catalogue with a 250\mic flux density fainter than 6.6 mJy), so
that sufficient unique positions can be generated. Inspecting the
ratio of the extracted and simulated input fluxes -- Figure
\ref{ext_true_grid_plots} -- suggests that the majority of the
flux--enhancement seen in Figure \ref{ext_true_plots} is due to 
blended sources, rather than boosting due to noise. However, the
PSF--weighted ratio of the brightest to second brightest simulated input source
contributing to each source in the extracted catalogue (Figure
\ref{beam_hists}) appears to contradict this; it shows that, even at
500\mic, blending with this second source would not increase the
extracted flux density by the amount seen. The solution to this
apparent contradiction becomes clear when the PSF--weighted ratio of the contribution from
all the simulated input sources within a beam to the flux density of the simulated input
match is considered instead (Figure \ref{beam_hists_total}). Here $\sim$27\% of 500\mic $> 5\sigma$ extracted sources have sufficient
simulated input sources available to boost their flux densities by a factor of 2
or more when their contributions are combined, even though their
individual effect is small. Figure \ref{beam_hists_total} also shows
that this confusion becomes negligible for $>10\sigma$ sources. This
is in broad agreement with \citet{chapin10} who find that the sub--mm
peaks they detect using a survey with larger beams, but of similar depth to H--ATLAS, generally consist of a blend of several sources.
Future versions of MADX will include a deblending step which should
reduce this effect. It should be noted that a mean sky--background of 6.8 mJy, 5.8 mJy or 4.1
mJy at 250\mic, 350\mic and 500\mic respectively (determined from the
mean of the simulated input catalogue), is subtracted before the histograms are calculated, to
account for the background--subtraction carried out as part of the
source extraction process. 

As a check on the success of the correction factors in Table
\ref{corr_table}, they are applied to the full extracted catalogues
and the fractional flux density errors (after rejecting the points which
lie outside the 99.73rd percentile) are then calculated. 
As Figure \ref{flux_err} shows, these reduce with
increasing S/N, but, as with the positional errors
discussed previously, Eddington bias prevents this behaving exactly as
expected. Again, when plotted against the S/N from the simulated input catalogue
(Figure \ref{flux_err2_pss}) the difference is reduced.  

\begin{figure}
\centering
\subfloat[The differential source
  counts for the extracted, simulated input (true) and flux--corrected catalogues for
  the three bands. Note the discrepancy between the flux--corrected
  and simulated input catalogues at faint flux
  densities.]{\protect\label{diff_cnt_plots_ess} \includegraphics[scale=0.6, clip, trim=0mm 3mm 0mm 4mm]{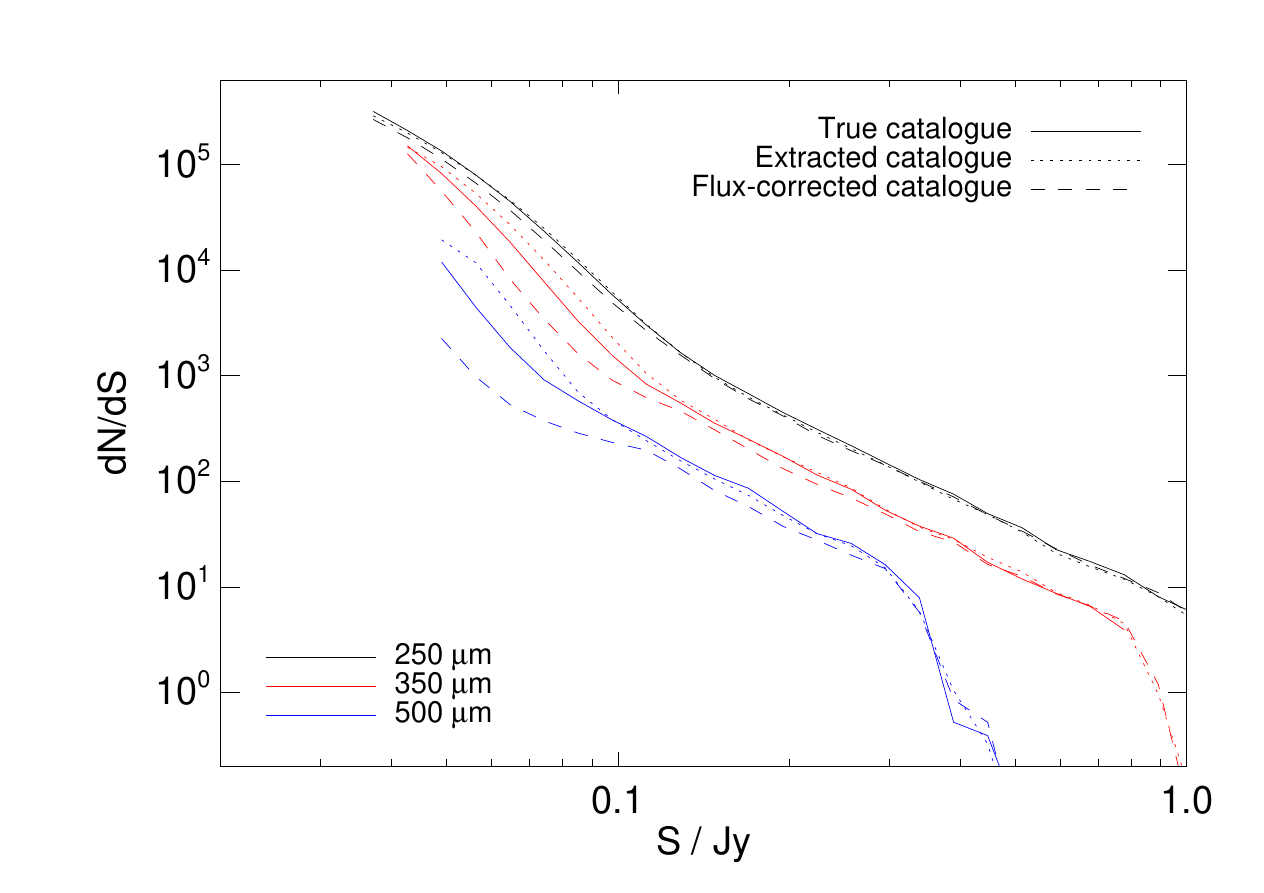}} \\
\subfloat[The surface density correction required at each band to
  correct for the catalogue incompleteness, determined from the ratio of the flux--corrected to
  simulated input differential source counts. The solid dots indicate the
  position of the average 5$\sigma$ limit in each
  band.]{\protect\label{comp_plots_ess}
  \includegraphics[scale=0.6, clip, trim=0mm 3mm 0mm 4mm]{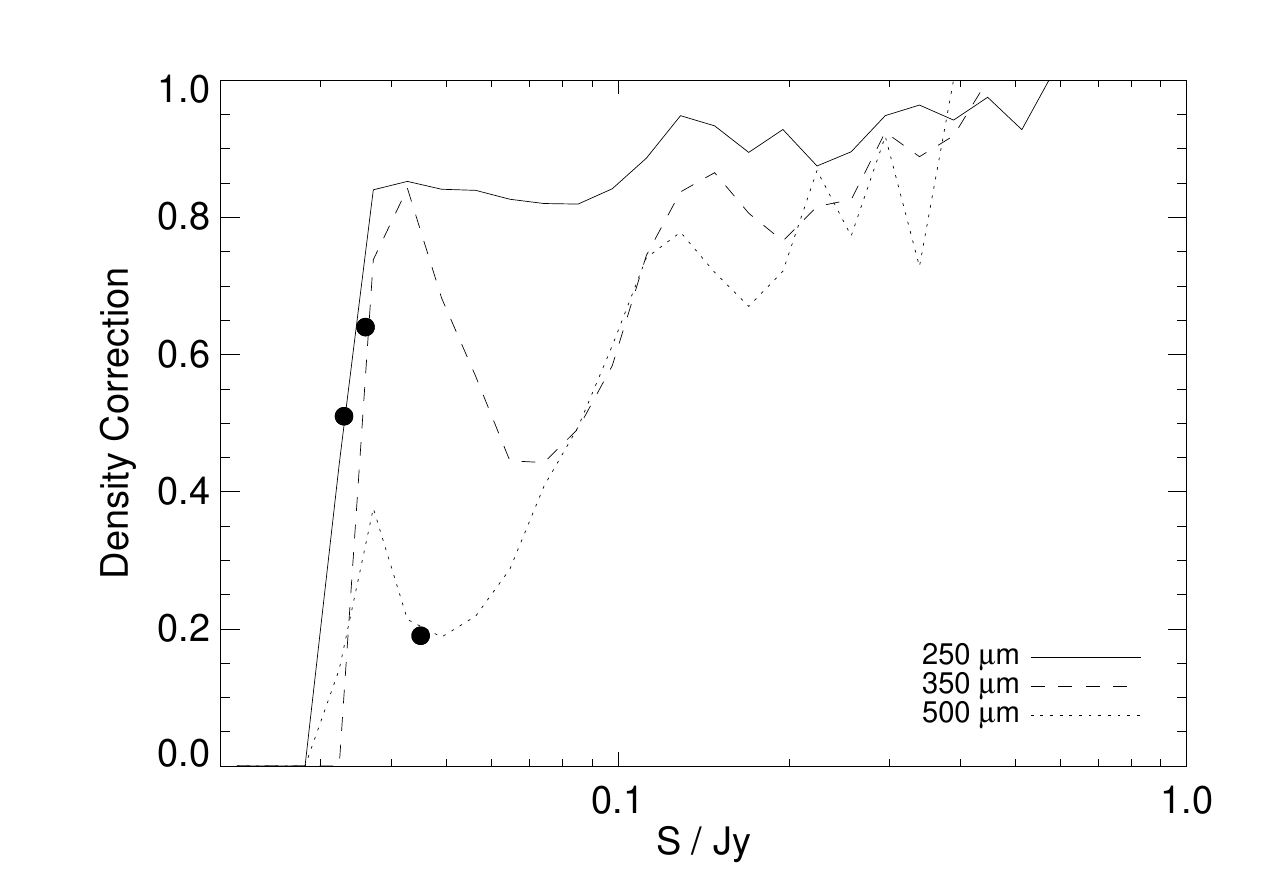}} \\ 
\subfloat[The integral source counts from
  the simulated input catalogue overplotted with the flux and
  surface--density corrected catalogue to demonstrate the success of
  these correction factors at recovering the simulated input
  values]{\protect\label{int_corr_ess} \includegraphics[scale=0.6, clip, trim=0mm 3mm 0mm 4mm]{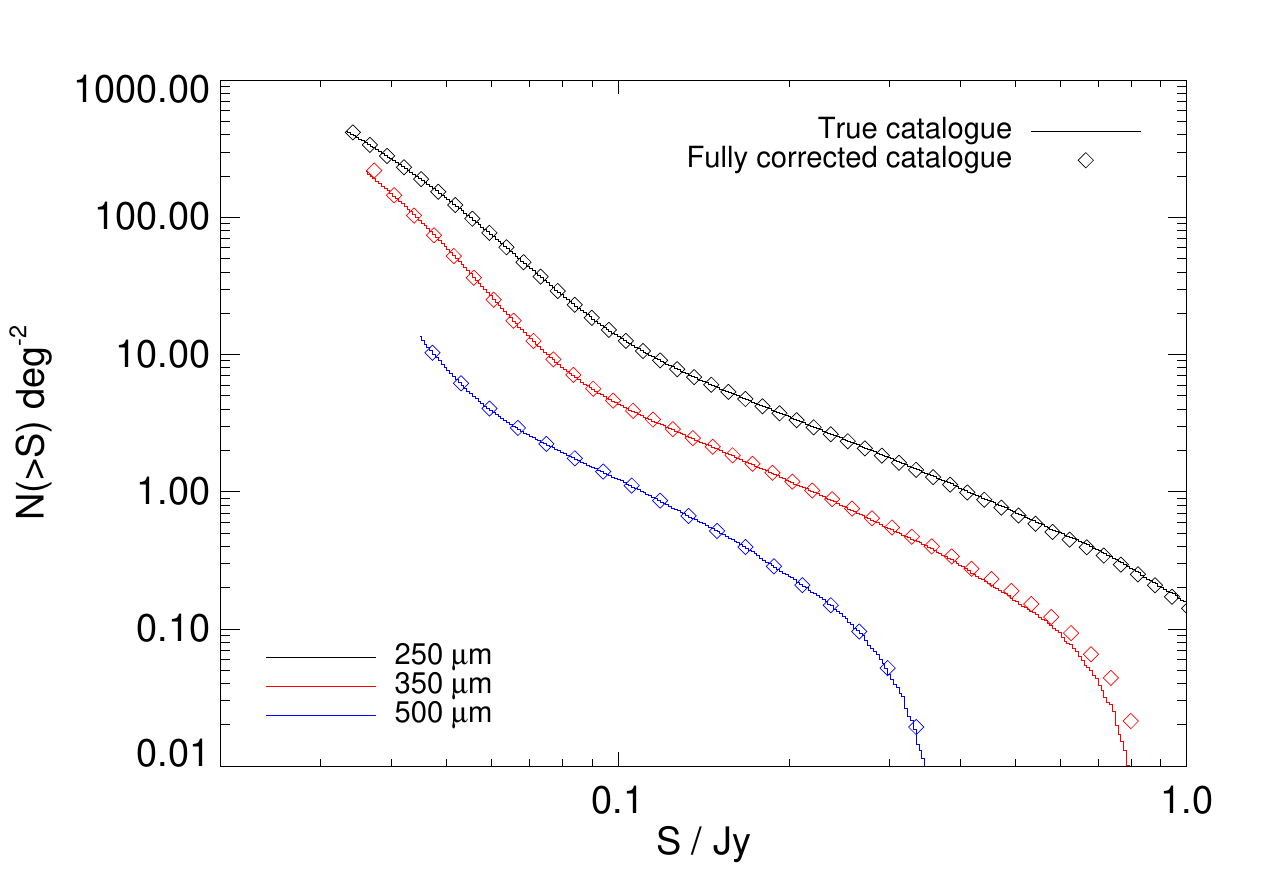}} 
\caption{\protect\label{aps_together_ess} Extended source simulations}
\end{figure}

\begin{figure}
\centering
\subfloat[The differential source counts for the extracted, simulated
  input (true) and flux--corrected catalogues for
  the three bands. Note the discrepancy between the flux--corrected
  and simulated input catalogues at faint flux
  densities.]{\protect\label{diff_cnt_plots_pss} \includegraphics[scale=0.6, clip, trim=0mm 3mm 0mm 4mm]{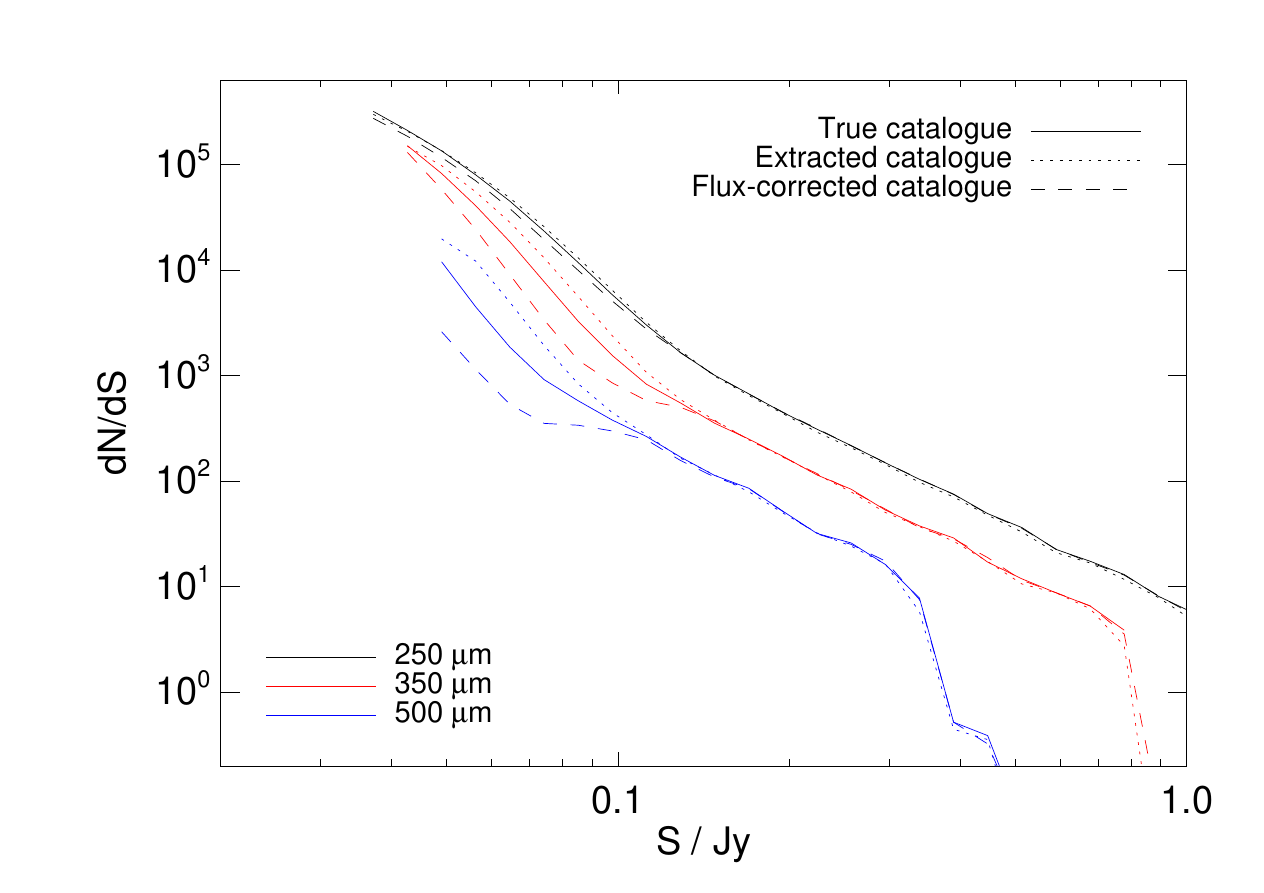}} \\
\subfloat[The surface density correction required at each band to
  correct for the catalogue incompleteness, determined from the ratio of the flux--corrected to
  simulated input differential source counts. The solid dots indicate the
  position of the average 5$\sigma$ limit in each
  band.]{\protect\label{comp_plots_pss}
  \includegraphics[scale=0.6, clip, trim=0mm 3mm 0mm 4mm]{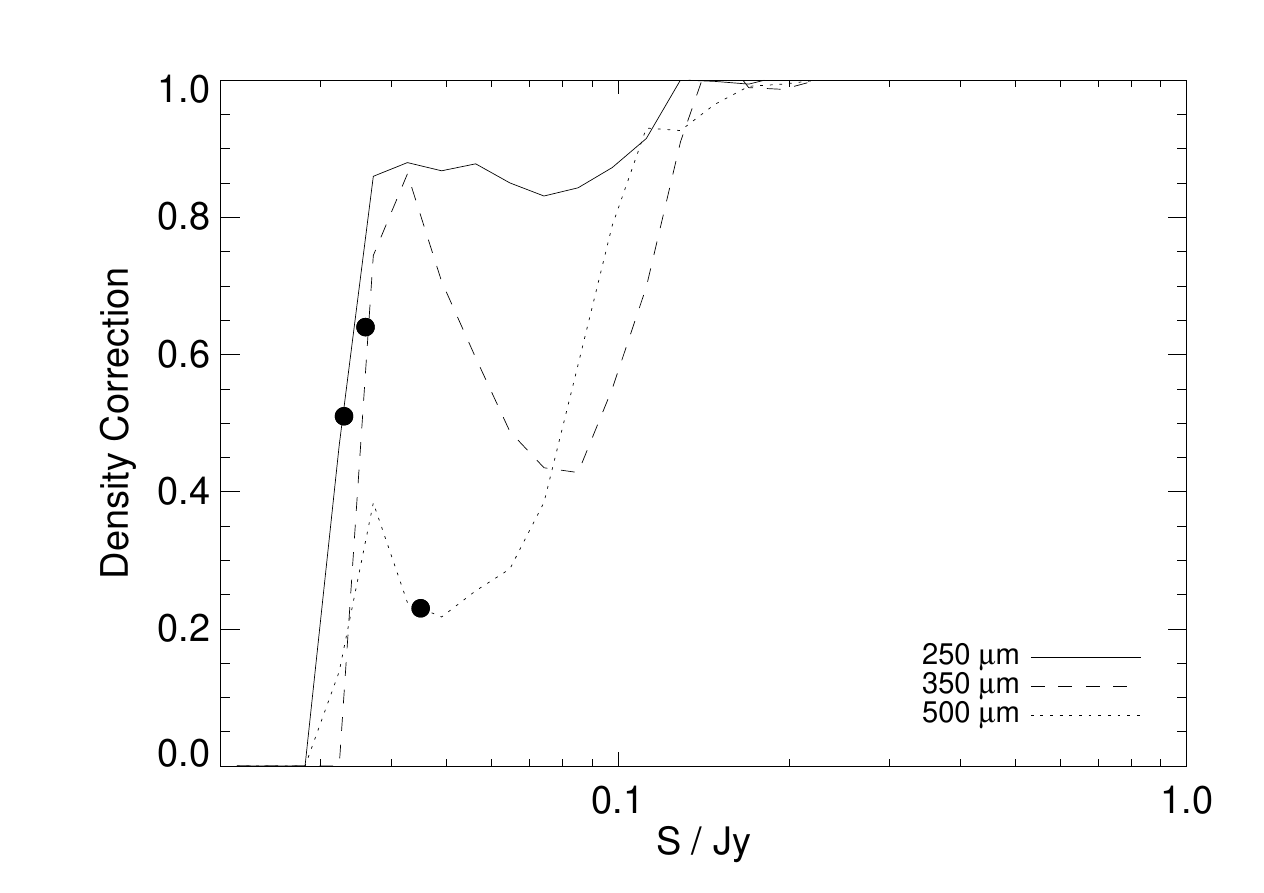}} \\  
\subfloat[The integral source counts from
  the simulated input catalogue overplotted with the flux and
  surface--density corrected catalogue to demonstrate the success of
  these correction factors at recovering the simulated input
  values]{\protect\label{int_corr_pss} \includegraphics[scale=0.6, clip, trim=0mm 3mm 0mm 4mm]{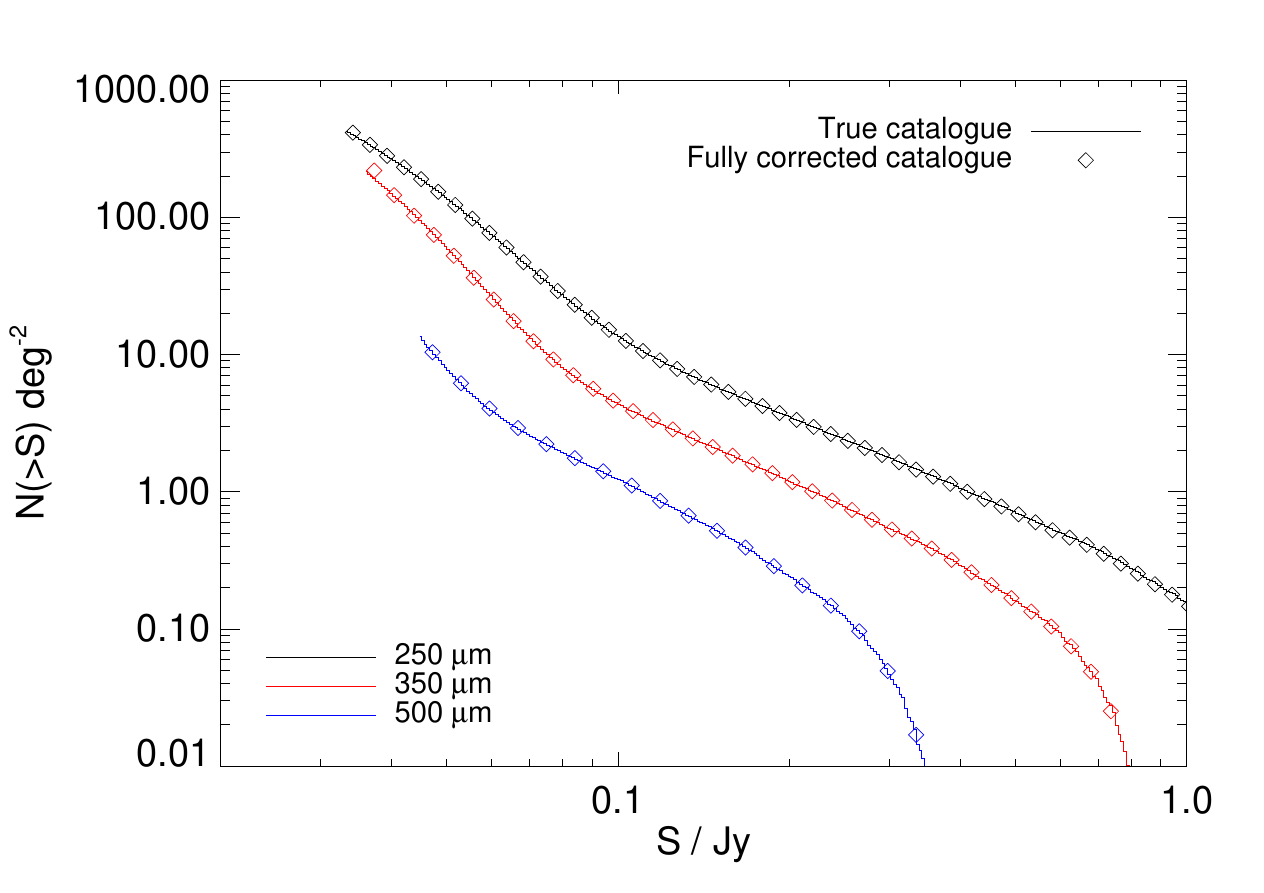}} 
\caption{\protect\label{aps_together_pss} Point source simulations}
\end{figure}

As well as the flux correction factors, we also need to completeness
of the detected catalogues, especially at faint 350 and 500\mic flux densities; this is clearly seen in Figures
\ref{diff_cnt_plots_ess} and \ref{diff_cnt_plots_pss}  which compare the differential source counts for
the extracted, simulated input and flux--corrected catalogues. The
lower counts are due to the failure to detect some fraction of faint sources
because of random noise fluctuations in the simulated maps or source blending. This incompleteness can be quantified by simply taking the ratio
of the flux--corrected to simulated input differential counts, to give a 
source--surface--density correction. Note that this is not appropriate for correcting the flux densities of
individual sources, but rather it can be applied when making
statistical analyses of the catalogue as a whole. This correction is shown in Figures \ref{comp_plots_ess} and \ref{comp_plots_pss}, and also given
as an additional correction factor in Table \ref{comp_table}. Figures
\ref{int_corr_ess} and \ref{int_corr_pss}  demonstrate the success of the density correction when applied to the integral source counts.

There is one further factor that can affect the extracted catalogue -- contamination from
spurious sources. The expected number of $\geq 5\sigma$ random noise
peaks present in the 250\mic map area is only $\sim$0.05, so this should be
negligible in the SDP catalogue. Contamination from fainter sources which are boosted or blended is accounted for in the flux correction factors. 

It should be noted that an alternative approach to correcting the SDP
H--ATLAS catalogue was adopted in \citet{clements}. In this case
corrections were determined from the ratio of extracted to simulated input
integral source counts. This combines the effects of incompleteness
and flux boosting, and is appropriate for recovering the correct
source counts, but not for correcting individual catalogue sources.  

\section{Concluding remarks}

This paper has presented the SDP catalogue for the first observations
of the H--ATLAS survey, along with a description of the simulations
created to determine the factors needed to correct it for
the combined effects of incompleteness, flux--boosting and source blending. 
The main results of this analysis are summarised below:
\begin{enumerate}

\item The extracted flux densities of 350\mic and 500\mic sources can be enhanced
  over their simulated input values, by factors of up to $\sim$2. This predominantly affects sources with $5<$ S/N $< 15$;

\item These enhancements are shown to be due to source blending, with  $\sim$27\% of $>5\sigma$ 500\mic sources having sufficient simulated input
  sources available within a beam to create a boosting of $\sim$2;

\item A combination of flux density and source--surface--density
  corrections are necessary to correct the extracted source counts
  for these factors.

\end{enumerate}

It is anticipated that future development of the MADX software will
incorporate subroutines to deal with both the effects of map
pixelization and source blending in the processing stage. 

MADX is not the only source extraction method being considered for the
H--ATLAS data, but time constraints mean that it has been used for the
SDP catalogue presented here. A comparison between different source
extraction algorithms is currently ongoing; these include
SUSSEXtractor developed by \citet{savage}, as well as the `matrix filter' method of
\citet{herranz} and the `Mexican Hat wavelet' method of \citet{mh1} and \citet{mh2}. The results of this comparison will be used to improve future H--ATLAS catalogues. 

This initial, uncorrected, catalogue will be available from
\verb1http://www.h-atlas.org1, though it is expected that as the data
processing steps are refined it will undergo future updates.  

\section*{Acknowledgments}

The {\it Herschel}-ATLAS is a project with {\it Herschel}, which is an ESA space
observatory with science instruments provided by European-led
Principal Investigator consortia and with important participation from
NASA. The H-ATLAS website is \verb1http://www.h-atlas.org/1. U.S. participants in {\it
  Herschel}--ATLAS acknowledge support provided by NASA through a
contract issued from JPL. The Italian group acknowledges partial financial support from ASI/INAF agreement n. I/009/10/0.

\clearpage

\begin{table*}
\centering
\begin{tabular}{c|c|c|c|c|c|c}
\hline
& \multicolumn{3}{c}{ESS} & \multicolumn{3}{c}{PSS} \\
Catalogue S/N &  FC$_{250\mu m}$ & FC$_{350\mu m}$ &  FC$_{500\mu m}$ &  FC$_{250\mu m}$ & FC$_{350\mu m}$ &  FC$_{500\mu m}$ \\
\hline
%ess                                                  %pss                                      
5.30 & 1.06 & 1.12 & 1.45 &       1.06 & 1.12 & 1.45 \\  
5.94 & 1.06 & 1.18 & 1.51 &       1.06 & 1.18 & 1.50 \\  
6.67 & 1.06 & 1.21 & 1.51 &       1.06 & 1.21 & 1.50 \\  
7.48 & 1.06 & 1.23 & 1.47 &       1.06 & 1.23 & 1.45 \\  
8.39 & 1.06 & 1.25 & 1.35 &       1.06 & 1.25 & 1.32 \\  
9.42 & 1.06 & 1.27 & 1.14 &       1.06 & 1.26 & 1.11 \\  
10.57 & 1.06 & 1.27 & 1.01 &         1.06 & 1.25 & 1.02 \\ 
11.86 & 1.05 & 1.23 & 1.00 &         1.05 & 1.20 & 1.00 \\ 
13.30 & 1.04 & 1.15 & 0.99 &         1.04 & 1.10 & 0.99 \\ 
14.93 & 1.02 & 1.04 & 0.98 &         1.02 & 1.01 & 0.99 \\ 
16.75 & 1.00 & 1.01 & 0.98 &         1.00 & 0.99 & 0.98 \\ 
18.79 & 0.98 & 1.00 & 0.98 &         0.98 & 0.99 & 0.99 \\ 
21.08 & 0.97 & 0.99 & 0.97 &         0.98 & 0.98 & 0.98 \\ 
23.66 & 0.96 & 0.99 & 0.97 &         0.97 & 0.98 & 0.98 \\ 
26.54 & 0.96 & 0.99 & 0.96 &         0.97 & 0.98 & 0.98 \\ 
29.78 & 0.96 & 0.98 & 0.96 &         0.97 & 0.97 & 0.98 \\ 
33.42 & 0.96 & 0.98 & 0.98 &         0.97 & 0.98 & 0.98 \\ 
37.49 & 0.95 & 0.98 & 0.97 &         0.97 & 0.97 & 0.99 \\ 
42.07 & 0.96 & 0.98 & 0.96 &         0.97 & 0.97 & 1.00 \\ 
47.20 & 0.95 & 0.97 & 0.99 &         0.97 & 0.97 & 0.98 \\ 
52.96 & 0.95 & 0.97 & 0.96 &         0.97 & 0.97 & 0.99 \\ 
59.43 & 0.96 & 0.94 & 0.95 &         0.97 & 0.97 & 0.97 \\ 
66.68 & 0.95 & 0.93 & -- &         0.96 & 0.97 & -- \\ 
74.81 & 0.95 & 0.92 & -- &         0.97 & 0.97 & -- \\ 
83.94 & 0.97 & 0.93 & -- &         0.97 & 0.97 & -- \\ 
94.18 & 0.97 & 0.94 & -- &         0.96 & 0.97 & -- \\ 
\hline
\end{tabular}
\caption{\protect\label{corr_table} The flux density correction factors (FC) at each SPIRE
  wavelength, as a function of S/N in the extracted catalogue,
  determined from the ratio of flux densities in the matched extracted
  and simulated input catalogues. To apply the correction at some catalogue flux
  density, $f_{\rm cat}$: $f_{\rm corr } = f_{\rm cat} / FC$, though
  note that the density correction given in Table \ref{comp_table}
  should also be applied as well.} 
\end{table*}

\begin{table*}
\centering
\begin{tabular}{c|c|c|c|c|c|c}
\hline
Corrected flux density & \multicolumn{3}{c}{ESS} & \multicolumn{3}{c}{PSS} \\
(Jy)  &  SC$_{250\mu m}$ &  SC$_{350\mu m}$ &  SC$_{500\mu m}$  &  SC$_{250\mu m}$ &  SC$_{350\mu m}$ &  SC$_{500\mu m}$\\
\hline

%ess                             %pss                          
0.0320 & 0.31 & -- & --  & 0.40 & -- & -- \\
0.0327 & 0.75 & -- & 0.11  & 0.79 & -- & 0.11 \\
0.0335 & 0.84 & -- & 0.41  & 0.85 & -- & 0.39 \\
0.0343 & 0.84 & -- & 0.49  & 0.85 & -- & 0.49 \\
0.0351 & 0.83 & -- & 0.48  & 0.85 & -- & 0.48 \\
0.0359 & 0.85 & 0.01 & 0.46  & 0.86 & 0.01 & 0.46 \\
0.0367 & 0.84 & 0.68 & 0.39  & 0.86 & 0.71 & 0.41 \\
0.0376 & 0.85 & 1.36 & 0.31  & 0.87 & 1.36 & 0.33 \\
0.0385 & 0.83 & 1.36 & 0.26  & 0.86 & 1.36 & 0.27 \\
0.0394 & 0.83 & 1.34 & 0.25  & 0.86 & 1.34 & 0.26 \\
0.0403 & 0.85 & 1.07 & 0.25  & 0.88 & 1.11 & 0.27 \\
0.0427 & 0.85 & 0.84 & 0.21  & 0.88 & 0.86 & 0.24 \\ 
0.0490 & 0.84 & 0.68 & 0.19  & 0.87 & 0.71 & 0.22 \\ 
0.0562 & 0.84 & 0.57 & 0.22  & 0.88 & 0.60 & 0.26 \\ 
0.0646 & 0.83 & 0.45 & 0.29  & 0.85 & 0.49 & 0.29 \\ 
0.0741 & 0.82 & 0.44 & 0.41  & 0.83 & 0.43 & 0.38 \\ 
0.0851 & 0.82 & 0.49 & 0.49  & 0.84 & 0.43 & 0.59 \\ 
0.0977 & 0.84 & 0.58 & 0.61  & 0.87 & 0.55 & 0.79 \\ 
0.1122 & 0.89 & 0.74 & 0.74  & 0.92 & 0.70 & 0.93 \\ 
0.1288 & 0.95 & 0.84 & 0.78  & 1.00 & 0.91 & 0.93 \\ 
0.1479 & 0.93 & 0.87 & 0.72  & 1.00 & 1.05 & 0.96 \\ 
0.1698 & 0.89 & 0.81 & 0.67  & 0.99 & 0.99 & 0.99 \\ 
0.1950 & 0.93 & 0.77 & 0.72  & 1.01 & 0.99 & 0.99 \\ 
0.2239 & 0.87 & 0.82 & 0.87  & 1.00 & 1.00 & 1.00 \\ 
0.2570 & 0.90 & 0.83 & 0.77  & 1.00 & 1.00 & 1.00 \\ 
0.2951 & 0.95 & 0.92 & 0.92  & 1.00 & 1.00 & 1.00 \\ 
0.3388 & 0.96 & 0.89 & 0.73  & 1.00 & 1.00 & 1.00 \\ 
0.3890 & 0.94 & 0.92 & 1.00  & 1.00 & 1.00 & 1.00 \\ 
0.4467 & 0.98 & 1.00 & 1.00  & 1.00 & 1.00 & 1.00 \\ 
0.5129 & 0.93 & 1.00 & 1.00  & 1.00 & 1.00 & 1.00 \\ 
0.5888 & 1.02 & 1.00 & 1.00  & 1.00 & 1.00 & 1.00 \\ 
0.6761 & 1.00 & 1.00 & 1.00  & 1.00 & 1.00 & 1.00 \\ 
0.7762 & 1.00 & 1.00 & 1.00  & 1.00 & 1.00 & 1.00 \\ 
0.8913 & 1.00 & 1.00 & 1.00  & 1.00 & 1.00 & 1.00 \\ 
1.0233 & 1.00 & 1.00 & 1.00  & 1.00 & 1.00 & 1.00 \\ 
\hline
\end{tabular}
\caption{\protect\label{comp_table} The surface density correction (SC) at each SPIRE
  wavelength as a function of corrected flux density,  determined from the ratio of the flux--corrected to simulated input
  differential counts.  To apply the correction at some corrected flux
  density, $f_{\rm corr}$: $f_{\rm corr\_final } = f_{\rm corr} /
  SC$. The corrected flux densities given are the central bin values.}
%\textcolor{red}{\it check this!}} 
\end{table*}

\bsp

\label{lastpage}

\end{document}